\documentclass[conference,compsoc]{IEEEtran}
\IEEEoverridecommandlockouts
\usepackage{cite}
\usepackage{amsmath,amssymb,amsfonts}
\usepackage{algorithmic}
\usepackage{graphicx}
\usepackage{textcomp}
\usepackage{xcolor}
\def\BibTeX{{\rm B\kern-.05em{\sc i\kern-.025em b}\kern-.08em
    T\kern-.1667em\lower.7ex\hbox{E}\kern-.125emX}}
\usepackage{booktabs} % for tables I think
\usepackage{dirtytalk}
\usepackage{multirow} % for tables
\usepackage{siunitx}
\usepackage[color=yellow]{todonotes}

\usepackage{lscape} % or pdflscape
\usepackage{rotating}

\usepackage[most]{tcolorbox}

\usepackage{paralist}
\begin{document}

\title{Self-Supervised Representations for Binary Program Clustering: From Empirical Study to Retrieval-Augmented Learning\\
}

% \author{
% \IEEEauthorblockN{{Martin Mocko}}
% \IEEEauthorblockA{\textit{Faculty of Information Technology} \\
% \textit{Brno University of Technology}\\
% Brno, Czech Republic \\
% martin.mocko@kinit.sk}
% \and
% \IEEEauthorblockN{Daniela Chud\'a}
% \IEEEauthorblockA{\textit{Faculty of Electrical Engineering and Information Technology} \\
% \textit{Slovak University of Technology}\\
% Bratislava, Slovakia \\
% daniela.chuda@stuba.sk}
% }

\author{
\IEEEauthorblockN{
Martin Mocko\IEEEauthorrefmark{1}\IEEEauthorrefmark{2} and
Daniela Chud\'a\IEEEauthorrefmark{3}\IEEEauthorrefmark{2}
}
\IEEEauthorblockA{\IEEEauthorrefmark{1}Faculty of Information Technology, Brno University of Technology, Brno, Czech Republic}
\IEEEauthorblockA{\IEEEauthorrefmark{2}Kempelen Institute of Intelligent Technologies (KInIT), Bratislava, Slovakia}
\IEEEauthorblockA{\IEEEauthorrefmark{3}Faculty of Electrical Engineering and Information Technology, Slovak University of Technology, Bratislava, Slovakia}
\IEEEauthorblockA{Email: martin.mocko@kinit.sk, daniela.chuda@stuba.sk}
}

% \and
% \and
% \and
% \and

\maketitle

\begin{abstract}
% The ever-increasing volume of sophisticated malware necessitates robust and automated methods for analysis, with malware clustering being an important way for identifying novel threats and threat families. While self-supervised learning (SSL) and contrastive learning (CL) have revolutionized computer vision and natural language processing, their application to tabular data, specifically in the malware domain - such as extracted malware features, remains challenging due to the domain-specific nature of existing techniques.

% The research is conducted in two phases: Phase 1 establishes a performance \say{ceiling} by adapting popular vision-based SSL models (BYOL, SimSiam, Barlow Twins, and VICReg) for tabular data using supervised pair generation. Phase 2 evaluates the effectiveness of purely unsupervised TRL methods, including VIME, SCARF, and SwitchTab, against strong traditional baselines such as PCA, Autoencoder, and UMAP. Using Homogeneity as the primary quality metric on the public Ember and Bodmas datasets, our results demonstrate that BYOL and SimSiam are the most effective SSL architectures for this domain, while Barlow Twins and VICReg significantly underperform. Crucially, we show that unsupervised TRL methods, particularly VIME, can outperform strong baselines and even approach the performance of fully supervised models. These findings establish a new state-of-the-art for binary program clustering and highlight the potential of self-supervised tabular learning to enhance automated malware analysis. Code will be made available.

Malware clustering is a critical task in cybersecurity that helps discover threats and analyze evolving malware families. While self-supervised learning (SSL) and tabular representation learning (TRL) have achieved breakthroughs in other domains, their application to binary program clustering (the task of clustering all incoming samples regardless of label) remains largely unexplored.
This study presents the first systematic investigation of SSL and TRL methods for binary program clustering, conducted in two phases on the public Ember and Bodmas datasets. In Phase 1, we establish a performance ceiling by adapting prominent vision-based SSL models (BYOL, SimSiam, Barlow Twins, VICReg) for tabular data with supervised pair generation, finding that BYOL and SimSiam achieve performance comparable to fully supervised models, while Barlow Twins and VICReg significantly underperform. In Phase 2, we evaluate purely unsupervised TRL methods against strong baselines (PCA, Autoencoder, UMAP), demonstrating that VIME establishes a new state of the art for binary program clustering. Informed by these findings, we propose VIME-R, a retrieval-augmented extension of VIME that replaces random marginal-distribution corruption with retrieval-based augmentation to generate more informative training pairs. VIME-R further improves upon VIME, achieving 2.7\%-5.8\% higher Homogeneity on both datasets. Our results highlight retrieval-augmented tabular representation learning as a promising direction for enhancing automated malware analysis. Code will be made available.

% Existing tabular SSL methods like VIME rely on global feature distributions for value imputation, which may introduce noise when applied to binary programs due to the behavioral differences between disparate malware families. To address this, we propose NIME, which constrains the imputation task to a sample's k-nearest neighbors, forcing the encoder to learn highly discriminative, localized structural representations. Moreover, we present a novel study on the applicability of well-established SSL techniques for malware clustering. We first establish the viability of these techniques by studying their performance in a supervised setting. We then move into the self-supervised learning scenario and empirically demonstrate that adapting tabular representation learning (TRL) methods significantly improves malware clustering state-of-the-art on public benchmark datasets, resulting in new state-of-the-art performance. Crucially, we show that the strategy for creating sample pairs in SSL and TRL is paramount to the subsequent quality of the learned representations. Our comprehensive analysis of the resulting cluster sets provide critical insights for applying representation learning in the malware field. Code will be made available.
\end{abstract}

\begin{IEEEkeywords}
% malware clustering, self-supervised learning, tabular representation learning, unsupervised learning, Ember, Bodmas

malware clustering, binary program clustering, self-supervised learning, tabular representation learning, retrieval-augmented learning
\end{IEEEkeywords}

\section{Introduction}
\label{sec:introduction}
More than 80 million new malware samples were created in 2024 alone\footnote{https://portal.av-atlas.org/malware/statistics}. Moreover, the average cost of recovery from a malware attack has risen from \$1.82 million in 2023 to \$2.73 million in 2024\footnote{https://www.sophos.com/en-us/press/press-releases/2024/04/ransomware-payments-increase-500-last-year-finds-sophos-state}. The trend of increasingly costly malware attacks motivates cybersecurity defenders to enhance their malware identification and prevention capabilities year after year. 

% https://spacelift.io/blog/malware-statistics

One task that can aid in the malware discovery and analysis process is malware clustering. It has the potential to identify new (potentially even zero-day) malware families \cite{jurevckova2024classification}, aid domain experts in malware analysis tasks\footnote{https://www.humansecurity.com/learn/blog/satori-threat-intelligence-disruption-}, and can aid in creating compressed representative datasets \cite{bayer2009scalable}, i.e., by sampling from the created clusters instead of using all collected samples. Moreover, binary program clustering - the task of clustering all incoming files, regardless of their potential label (i.e., malware, benign) was recently recognized to be a potential new way forward for clustering samples while still retaining high clustering quality \cite{mocko2025clustering}. 

\begin{figure}[!t]
    \centering
    % [width=0.75\columnwidth]
    \includegraphics[height=6.5cm, width=\columnwidth, keepaspectratio]{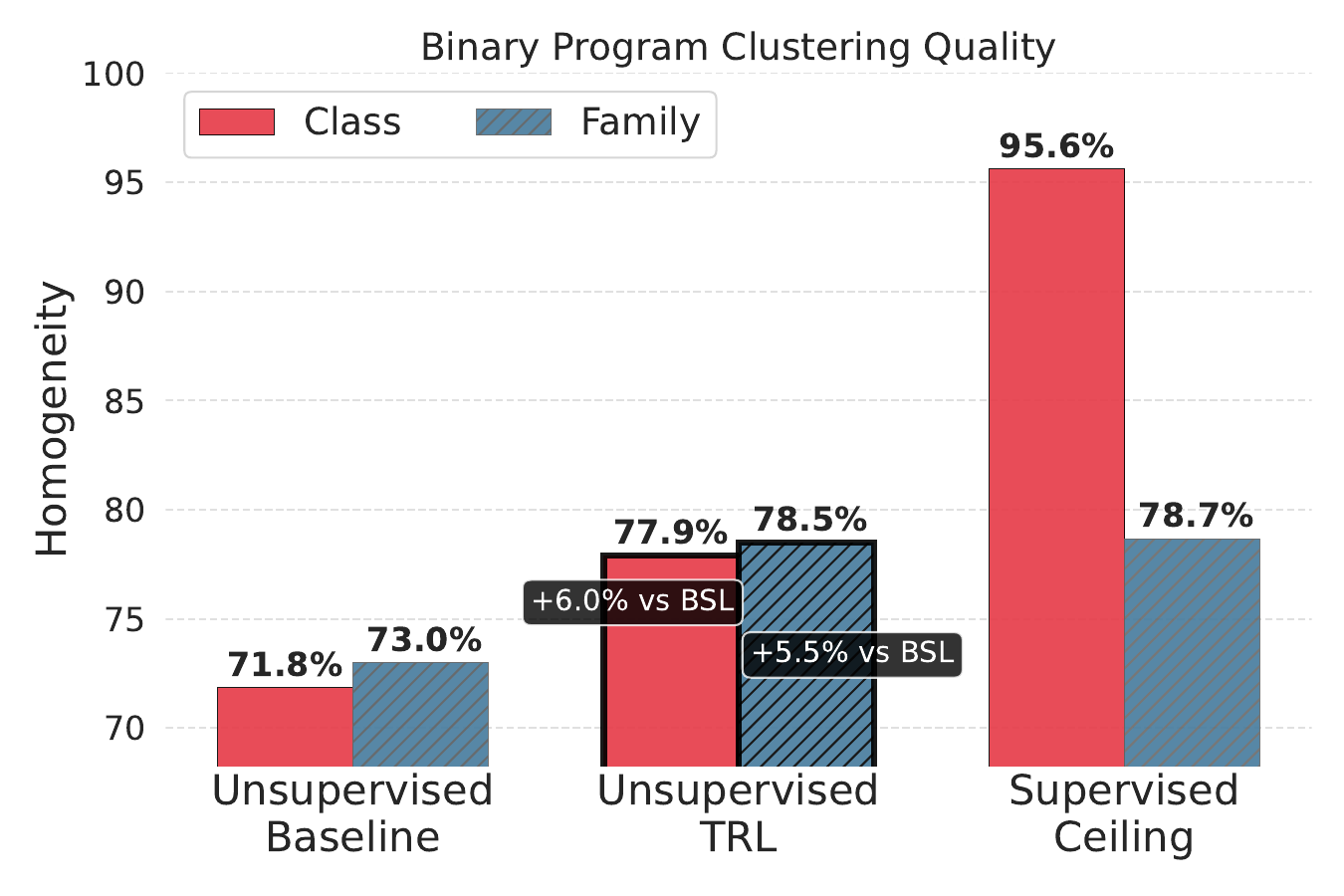} % Or filename.pdf/.png
    % \caption{An overview of the clustering quality achieved via the utilization of tabular representation learning methods for binary program clustering (averaged over both utilized datasets). The quality gains are measured against strong unsupervised baselines. The Figure shows Homogeneity computed on both class and family labels. The supervised ceiling highlights the gap between the classic (fully) supervised solution, supervised SSL solutions, and unsupervised solutions. Improvements over baseline (referred to as \protect\say{BSL}) are shown in absolute percentages.}
    \caption{Clustering quality of TRL methods for binary program clustering (averaged across datasets), measured by Homogeneity on class and family labels. Gains are shown as absolute percentages over strong unsupervised baselines (BSL). The 'Supervised Ceiling' illustrates the performance gap between the group of fully supervised models/supervised SSLs and unsupervised solutions. Results computed as mean of the Homogeneity results for top-3 best methods for each method category.}
    \label{fig:aggregated_results}
\end{figure}

However, to perform malware clustering efficiently, the input features (representations) presented to a clustering technique must be as representative and informative as possible. Such representations are often a result of a certain set of conditions, i.e.: a) fully unsupervised - no labels are available, b) semi-supervised - few labels are available, c) fully supervised - all labels are available. 

In the last few years, many self-supervised (SSL) and contrastive learning (CL) approaches have been shown to work very well in the computer vision and natural language processing (NLP) domains \cite{grill2020bootstrap,chen2021exploring,devlin2019bert}. However, the achievements in these domains do not automatically translate to the tabular data domain, specifically the malware domain. The main problem with transferring these methods is that they rely heavily on data augmentation to create pairs that serve as model inputs.

%These data augmentations are highly dependent on correlations which are often not present in the case of tabular data. 
These data augmentations aim to preserve semantic content while perturbing low-level (for example, image) statistics. Such augmentations are often not present in the case of tabular data.
% For example, in the case of computer vision tasks, they are local correlations in the pixel space. 
For NLP tasks, they are often sequential correlations (there is a certain sequential order in natural language text). The tabular domain does not contain these correlation advantages, and therefore creating data augmentations for tabular data is a much harder task.
Nevertheless, multiple tabular representation learning (TRL) approaches were proposed over the last few years that also seem to contribute to the state of the art in self-supervised tabular representation learning \cite{yoon2020vime,bahri2021scarf,wu2024switchtab,ucar2021subtab}. 

Our work explores the potential of self-supervised and tabular representation learning \emph{in an unsupervised context} to improve the quality of malware clustering, or rather, \emph{binary program clustering}. For our baseline, we adopt and replicate the unsupervised representation learning methods used in \cite{mocko2025clustering} as a strong baseline (as confirmed by our experiments) and aim to beat it. We divide our experiments into two phases. The first phase of our experiments investigates the ceiling performance of well-known SSL methods that we adapted for the tabular domain. The ceiling is tested using pair creation based on the sample label, i.e., the pairs are supervised. The second phase of the experiments explores whether the reported improvements in recent tabular representation learning methods will also manifest in the malware domain in an unsupervised context  (for binary program clustering).

For the conducted experiments, we utilize the Homogeneity metric as our primary indicator of clustering quality. Multiple other malware clustering works \cite{fang2019semi,wilkins2020cougar,jurevckova2024classification,jurevckova2025online,mocko2025clustering} used Homogeneity or Purity (a very similar metric) as their main indicator of clustering quality in the past as well. The primary motivation is that, for clustering to be useful in the malware domain, it should first be able to create homogeneous clusters (of malware families). In addition, we report Homogeneity for both the class label (malware/benign) and the \emph{malware family} label, emphasizing the differences between the two evaluation measures. 

% Our results show that the ceiling of SSL methods (at least some of them) is high and comparable to that of more typical supervised approaches, such as an MLP trained to recognize individual family labels. A general overview of the average results can be seen in Figure \ref{fig:aggregated_results}. Moreover, we find that some SSL methods do not seem to perform as well as others. 
% From purely unsupervised learning experiments, we find that self-supervised tabular learning approaches can beat the previous state of the art and thus bring meaningful improvements to unsupervised (or self-supervised) tabular learning for binary program clustering. The classical SSL methods, however, suffer when they do not receive very specific sample pairs from the same malware family, rendering their use much more difficult. 
% Finally, we analyze the generated clusters and briefly describe the main differences between the unsupervised baseline, the best unsupervised TRL solution, and supervised clusters.

Our results demonstrate that self-supervised tabular learning can significantly outperform existing state-of-the-art approaches in binary program clustering. Specifically, we find that the performance \emph{ceiling} of certain (originally vision-based) SSL approaches (when trained using supervised pairs) is comparable to fully supervised models, such as Multi-Layer Perceptrons (MLPs) trained on individual family labels (see Figure \ref{fig:aggregated_results}). While we observe that classical SSL methods struggle without highly specific, family-aligned sample pairs, our best-performing \emph{fully unsupervised} TRL solution (VIME-R) bridges this gap, offering a meaningful improvement over traditional unsupervised baselines and achieving \emph{new state-of-the-art results} in binary program clustering on both Ember and Bodmas datasets, beating even vanilla VIME in the process. 

The contribution of our work is multi-fold and can be summarized as follows:
% \begin{compactitem}
%     \item Our study is the first... 
%     \item We demonstrate that... [cite: 502]
% \end{compactitem}

\begin{compactitem}
    \item Our study is the first to investigate the applicability of self-supervised learning (SSL) and tabular representation learning (TRL) specifically for the task of binary program clustering (or, malware clustering, in general);
    % \item We achieve new state-of-the-art results for malware clustering (more specifically, binary program clustering) in terms of Homogeneity on two public malware benchmark datasets;
    \item We demonstrate that unsupervised TRL methods, particularly VIME, can outperform strong traditional unsupervised baselines like PCA and Autoencoders, setting new state-of-the-art results for binary program clustering on the public Ember and Bodmas datasets;

    % \item Through a two-phase experimental approach, we establish a \say{performance ceiling} for clustering binary program representations based on pair learning, via the utilization of traditional SSL methods and using supervised pair generation, providing a benchmark for what these models can achieve in the malware domain when ideal positive pairs are available;

    \item By adapting foundational SSL methods via supervised pair generation, we establish a performance 'ceiling' for binary program clustering. This benchmark enables a direct comparative analysis for novel data augmentation strategies, providing a baseline for their effectiveness in the malware domain;
    
    % \item We provide a comparative analysis that reveals that unsupervised TRL methods, such as VIME, achieve representational power by creating more specialized clusters for certain malware families. This behavior differs from the consolidation seen in fully supervised counterparts;

    \item We propose a novel retrieval-augmented corruption strategy for VIME, called VIME-R, that significantly improves the VIME method on the task of binary program clustering and achieves state-of-the-art results, beating all other tested unsupervised models.

% \item We show that not all popular SSL methods are created equal, at least for tabular malware data, and that models like BYOL and SimSiam are more suitable to use for the binary program clustering task than BarlowTwins and VICReg;

% \item To the best of our knowledge, this is the first work to adapt SSL models originating from the vision domain (BYOL, SimSiam, Barlow Twins, and VICReg) for use on tabular data;
    
\end{compactitem}

The rest of the paper is organized as follows. Section \ref{sec:relwork} explores the work related to this research. Section \ref{sec:method} introduces our approach undertaken to explore the possibilities of applying self-supervised and tabular representation learning approaches for the malware clustering task. In Section \ref{sec:experiments}, results from the performed experiments are presented. It also introduces our proposed modification to VIME, called VIME-R. Section \ref{sec:discussion} draws conclusions from the experiments and discusses the broader implications of the work. Finally, section \ref{sec:conclusion} concludes the work with final statements and takeaways.

% \todo[inline]{Perhaps mention in the intro that our augmentations/corruptions are not specific to the malware domain, but general }

\section{Related Work}
\label{sec:relwork}

\subsection{Self-Supervised and Contrastive Learning in the Malware Domain}
% here I should show some recent progress in the malware clustering domain 
% as well as show that there are some papers which explored the usage of SSL/CL at least for malware detection; if I can find something for malware clustering, include it as well

% Malware clustering is the less-explored sibling of malware detection and malware family classification that often falls behind in terms of research. 

% I need to publish the updated result of clustering using AE, UMAP, PCA on EMBER !!!!!!!

% mention the term "zero-day" in the paper, perhaps also "intrusion" or something like that

% I guess specify what are the results achieved in the past for malware clustering, maybe also on what datasets

% also specify the latest result from us -- the specific Homogeneities

% talk about the use of SSL / CL / TRL --> first, ideally for malware clustering, but then also for malware detection perhaps ??

% perhaps it would be good to just show the work that was done in the last few years

Despite numerous recent works in malware clustering \cite{ali2020scalable,wilkins2020cougar,jurevckova2024classification,jurevckova2025online,macaskill2021scaling,mishra2025cluster,mocko2025clustering}, self-supervised, contrastive, and tabular representation learning remain under-explored. Mocko et al. \cite{mocko2025clustering} recently established baselines by comparing standard clustering and dimensionality reduction (PCA, Autoencoder, UMAP) on public benchmarks. Due to the scarcity of malware-specific clustering literature using these techniques, we include relevant detection and classification studies. Some utilize non-tabular graph structures (e.g., CFGs) \cite{gao2023malware,su2024graph}, while others transform binaries into grayscale or RGB images to apply vision-based augmentations \cite{ismail2024malssl,wang2025self} or ViT-style masking \cite{wang2024malsort}.

% \cite{gao2023malware,gao2022unsupervised,su2024graph}
% \cite{jia2023imcscl,ismail2024malssl,wang2025self}
% \cite{wang2024malsort,seneviratne2022self}

Other approaches employ specialized pre-training: Carter et al. \cite{carter2025contrastbert} use Linux syscalls in a BERT-like contrastive framework, while Wang et al. \cite{wang2023bibe} and Trizna et al. \cite{trizna2024nebula} apply masked language modeling to static Ember data and dynamic reports, respectively. To address concept drift, CADE \cite{yang2021cade} combines autoencoders with supervised contrastive loss, and EVOLIoT \cite{dib2022evoliot} utilizes BERT with dropout-based positive pairing.

\subsection{Self-Supervised Learning Methods}

This subsection introduces impactful non-contrastive self-supervised learning (SSL) methods originally designed for computer vision. All of the methods utilize some form of a Siamese neural network architecture. Additionally, all of the methods also learn based solely on positive sample pairs. A primary challenge in non-contrastive SSL is avoiding "collapsing" solutions where the network produces identical outputs. BYOL \cite{grill2020bootstrap} addresses this using a momentum-based \say{teacher} network to bootstrap latent representations, while SimSiam \cite{chen2021exploring} simplifies this by using a stop-gradient operation instead of a momentum encoder. Alternatively, Barlow Twins \cite{zbontar2021barlow} prevents collapse by minimizing the cross-correlation matrix of distorted samples, and VICReg \cite{bardes2021vicreg} introduces explicit Variance, Invariance, and Covariance terms into the loss function to ensure architectural independence. While these methods achieve state-of-the-art results using image augmentations, their efficacy on tabular malware data remains an open question.

\subsection{Tabular Representation Learning Methods}
Tabular representation learning was solidified as a research area only a few years ago, with the 2022 workshop at NeurIPS titled the \say{First Table Representation Learning (TRL) Workshop}\footnote{https://neurips.cc/virtual/2022/workshop/49995}. Nevertheless, several TRL methods have emerged that gained popularity. 

Modern tabular representation learning often relies on feature-level corruption. VIME \cite{yoon2020vime} introduces this by replacing feature subsets with values from the empirical marginal distribution, using a \say{mask estimator} and reconstruction loss to learn representations. SCARF \cite{bahri2021scarf} adopts VIME’s corruption technique but replaces the reconstruction objective with a contrastive InfoNCE loss, reportedly yielding superior results. Most recently, SwitchTab \cite{wu2024switchtab} utilizes an asymmetric autoencoder to decouple mutual and salient features between sample pairs, reportedly outperforming VIME by better isolating unique sample characteristics.

To conclude the section, our focus is on improving malware clustering. Given the successes SSL and TRL models have achieved (admittedly in other domains), we consider them a potential way to improve the quality of malware clustering. Because the introduced SSL and TRL approaches may appear very similar, one could speculate that it is sufficient to utilize only TRL methods, as they are specifically aimed towards tabular data. However, it is important to emphasize that the situation is not as clear-cut. While TRL methods are designed for tabular data, they are not plug-and-play, as one cannot design their own sample-pair creation strategies for them. The only methods that are independent of the pair-creation strategies (beyond requiring positive pair inputs) are the SSL methods: BYOL, SimSiam, BarlowTwins, and VICReg. These methods support custom data augmentation (pair creation) strategies. Therefore, future improvements via custom augmentation techniques designed specifically for the malware domain could favor SSL methods over TRL methods. For TRL methods such as VIME, SCARF, and SwitchTab, the pair creation (or, rather, pair corruption) strategies are strictly defined. Therefore, tampering with them means altering the core of what makes the methods work so well. 

Moreover, the research conducted using SSL and TRL methods to improve malware clustering is demonstrably lacking. We are not aware of any malware clustering works that focus on improving clustering via SSL or TRL methods. Even the use of these methods on tabular data, even in the field of malware detection, has so far been scarce. Finally, we were unable to find any works that used most of the proposed SSL methods, even for tabular data in general (across other domains). All of the aforementioned facts serve as motivation for our work.

% To conclude the section, we are not aware of any malware clustering works that focus on improving clustering via self-supervised, contrastive or tabular representation learning methods. Moreover, the usage of these methods on tabular data even for the field of malware detection has so far been scarce. Finally, we were not able to find any work utilizing the proposed SSL methods even for tabular data in general (in other domains). Therefore, our work is clearly motivated by the lack of research in the malware and tabular data space.

%%% Closing statements
% there are no -malware clustering- works that do the kind of experiments which we are about to do
% for supervised malware tasks, a lot of works utilized graph, image or sequential modalities (API/system calls or relatively free-form text) .. this is also mainly due to easier augmentations ... for tabular modality there are very few papers 
% what about tabular datasets in general? has this kind of study been done for tabular data in general ? 

\section{Clustering Malware and Benign Programs using TRL and SSL Representations}
\label{sec:method}
First, we reiterate that the ultimate goal of this work is to investigate the potential of increasing the quality of malware clustering (specifically, binary program clustering). To achieve this, we aim to improve the underlying representation that the clustering model operates on. Recall that self-supervised learning and tabular representation learning have both made contributions regarding the state-of-the-art in representation learning, as mentioned in Section \ref{sec:relwork}. Therefore, we investigate whether the utilization of self-supervised and tabular representation learning methods has the potential to improve the quality of malware clustering. 
% First, we aim to establish the ceiling of these methods and then we investigate how close is it possible to get to these ceilings using purely unsupervised methods.

More precisely, we consider the following two research questions:
\begin{compactitem}
    \item \textbf{RQ1:} Do all of the proposed popular SSL methods achieve a similar level of Homogeneity when utilizing \emph{supervised pair generation}? 
    % Are any of them worse than unsupervised methods ? 
    \item \textbf{RQ2:} Is it possible to beat a strong unsupervised baseline and improve binary program clustering quality by utilizing self-supervised or tabular representation learning methods? 
    % \item \textbf{RQ3:} How does the performance of popular SSL methods change when, instead of using supervised pair generation, we utilize unsupervised pair generation?  
\end{compactitem}
Based on the two research questions we divide the research into two phases - Phase 1 and Phase 2. The methodologies for Phase 1 and Phase 2 are the same in terms of the utilized datasets, preprocessing aproach, and baselines. Afterwards, the methodologies of Phase 1 and Phase 2 diverge. 

In Phase 1, we aim to experimentally measure the potential \emph{ceiling} of the devised popular self-supervised learning methods. 
More broadly, this performance level could constitute the ceiling for all methods that learn representations based on supervised positive pairs. 
% We speculate that this could potentially also be the ceiling of learnable representations based on the supervised positive pairs that the SSL models receive as their input. 
These SSL methods heavily rely on image augmentations and operate solely on positive pairs. In the malware domain, augmentation techniques to synthesize new malware/benign sample pairs are scarce. Moreover, they are often relatively simple (see \cite{demetrio2021adversarial}) and, more importantly, defined only for PE samples. This makes it impossible for researchers without access to the original binaries to perform the augmentations and observe the effect on the resulting vectorized features (which, arguably, can sometimes even be negligible). Therefore, to fully measure the ceiling of SSL methods, we use labeled data to generate positive sample pairs.

\begin{figure}[h!]
    \centering
    \includegraphics[width=0.75\columnwidth]{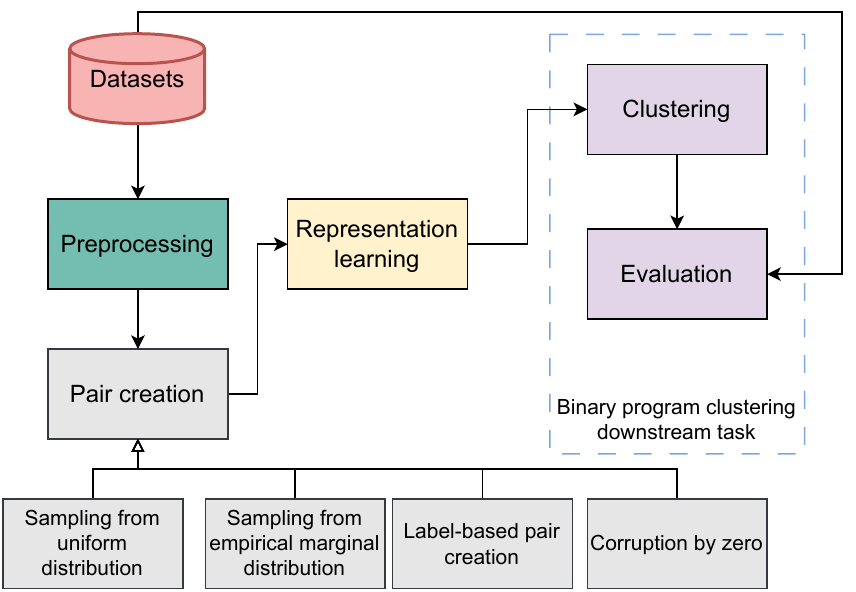} % Or filename.pdf/.png
    \caption{The general process of our methodology. Different representation learning methods are used for Phase 1 and Phase 2. Four distinct methods of pair creation are utilized overall. This is mainly due to the corruptions defined for VIME, SCARF, and SwitchTab.}
    \label{fig:methodology}
\end{figure}

Phase 2, on the other hand, operates solely in the unsupervised regime. Therefore, no class (i.e., malware/benign) or malware family labels are used at any point in the model training process. The two phases of our experiments use the same clustering approach and the same evaluation method. We reiterate that our specific malware clustering scenario is the \emph{binary program clustering} scenario, where we cluster all the (incoming) samples, regardless of their label. Thus, we do not rely on a pre-filtering step that removes benign programs from the dataset. A high-level view of our methodological framework is shown in Figure \ref{fig:methodology}.

% pridat definiciu klastering

% pridat definiciu binary program klastering a zdoraznit ze toto je nas kontext

\subsection{Datasets}
% \textbf{Ember} is the first real large public malware benchmark dataset \cite{anderson2018ember}. It is based on Windows PE files and contains features extracted from static analysis. Ember was published in two different \say{year} editions, 2017 and 2018, each with a different sample composition. Moreover, it was introduced with two feature sets - v1 and v2. For our experiments, we utilize Ember v2 2018. It is composed of \num{600000} training examples and \num{200000} test set examples. It has \num{2381} features in total and spans approximately \num{3000} AVclass families \cite{sebastian2016avclass}. Ember created the de facto PE file feature standard in the form of Ember features. Along with the features, the Ember authors also published code to extract them from PE file samples. 
\textbf{Ember} is a de facto standard for Windows PE static analysis \cite{anderson2018ember}. Ember consists of features extracted from raw binaries. This study utilizes the Ember v2 2018 edition, which comprises \num{600000} train and \num{200000} test samples across approximately \num{3000} AVclass \cite{sebastian2016avclass} families. Each sample is represented by a \num{2381}-dimensional feature vector. The dataset includes the original feature extraction source code to ensure reproducibility.

\textbf{Bodmas}, released in 2021, improves upon existing benchmarks by providing malware samples, expert-curated family labels, and temporal metadata (first-seen timestamps) \cite{yang2021bodmas}. It utilizes the full Ember feature set, ensuring compatibility between the two datasets. While both Ember and Bodmas rely on static features, which may offer a constrained view of PE behavior, they remain the standard for ensuring experimental repeatability and facilitating cross-study comparisons.  Summary information about the datasets is presented in Table \ref{tab:ds_info}.

% \textbf{Bodmas} \cite{yang2021bodmas} is a newer public malware benchmark dataset published in 2021. It improves on the malware dataset landscape by: 1) making malware samples available, 2) including timestamp information about when the malware was first seen, and 3) providing malware family labels curated by experts. Bodmas utilizes the full set of Ember features. Summary information about the datasets is presented in Table \ref{tab:ds_info}. For more information about Ember features and its categories, the reader is referred to the original Ember publication \cite{anderson2018ember}. We acknowledge that both datasets contain only static features, which may provide only a limited description of the PE samples. However, these datasets are popular public malware benchmark datasets that ensure experiment repeatability and allow comparison with other studies.

\begin{table}[]
\centering
\caption{Basic statistics about the datasets utilized for this study. M/B ratio represents the ratio between malware and benign samples in the respective dataset.}
\label{tab:ds_info}
\begin{tabular}{lrrrrr}
\toprule
\textbf{Dataset}   \hspace{-1em}    & \textbf{   \# features} & \hspace{0.1cm} \textbf{     Dataset size} & \textbf{M/B ratio}  & \textbf{   \# families} \\ \midrule \\[-10pt]
Bodmas                & \num{2381} & \num{134435} total  &    43:57                                             & \num{581}          \\[3pt]
Ember v2            & \num{2381} & \begin{tabular}[c]{@{}c@{}}\num{600000} train\\ \num{200000} test\end{tabular} &    50:50  & $\sim$\num{3000} \\\bottomrule
\end{tabular}

\end{table}

% \begin{table}[]
%     \centering
%     \caption{Description of the Ember features that can be found both in the Ember and Bodmas dataset. The vector is composed of \num{2381} features which are divided into nine categories. }
%     \label{tab:ember_features}
%     \begin{tabular}{c c}
%     \toprule
%       \textbf{Feature position}   & \textbf{Feature category} \\
%     \midrule 
%      0 - 255    & Byte histogram \\
%      256 - 511 & Byte entropy histogram \\
%      512 - 615 & String extractor \\
%      616 - 625 &  General file info \\
%      626 - 687    & Header file info \\
%      688 - 942    & Section information \\
%      943 - 2222    &  Imports information \\
%      2223 - 2350    & Exports information \\
%      2351 - 2380    & Data directories information \\
%      \bottomrule 
%     \end{tabular}
% \end{table}

\subsection{Preprocessing \& Baseline}
\textbf{Preprocessing}\quad With regards to the preprocessing steps, we use the approach utilized by \cite{mocko2025clustering}, who established the state-of-the-art results in malware clustering on Bodmas and Ember datasets. This preprocessing approach is undertaken in order to ensure comparability of results between these two works. Thus, we first utilize scalers from the \texttt{scikit-learn}\footnote{https://scikit-learn.org\label{fn:sklearn}} preprocessing module in the following order: RobustScaler, StandardScaler, and MinMaxScaler. Furthermore, we remove the same set of Ember features (for each dataset) identified as harming representation learning in \cite{mocko2025clustering}, which we gathered from their GitHub repository\footnote{https://github.com/kinit-sk/clustering-ares-2025}. This leaves us with 2235 Ember features for each dataset.

\textbf{Baseline}\quad For both of the utilized datasets, we adopt the unsupervised baselines (PCA using \texttt{scikit-learn}\footref{fn:sklearn}, Autoencoder using \texttt{PyTorch}\footnote{https://pytorch.org/}, and UMAP using \texttt{cuML}\footnote{https://docs.rapids.ai/api/cuml/stable/}) from \cite{mocko2025clustering}. Similarly to \cite{mocko2025clustering}, we also set the number of components (dimension size after dimensionality reduction) to 10. In the case of UMAP, we set the \emph{n\_neighbors} parameter to \num{20}, \emph{min\_dist}  to \num{0.1} and \emph{init} to spectral. 
The baselines that we utilize are strong unsupervised baselines, as will be shown later in the experimental results. Each baseline method is run 5 times to obtain robust metric information during evaluation.

        % n_neighbors=20,            # Slightly higher neighbors helps global structure
        % min_dist=0.1,
        % init='spectral',
        % n_epochs=1000,             # Gives the GPU time to find the true mathematical minimum
        % negative_sample_rate=15,   # Forces sharper cluster boundaries
        % build_algo="nn_descent",   # Use NN-descent but make it strict
        % build_kwds={
        %     "nnd_graph_degree": 64, 
        %     "nnd_max_iterations": 30

% Note that the results reproduced on these baselines are not exactly the same, since we use \texttt{PyTorch} for our experiments, whereas Mocko et al. \cite{mocko2025clustering} used \texttt{TensorFlow}. PCA also has low variability, as does UMAP. Nevertheless, the results for the baselines are very similar to \cite{mocko2025clustering}. 

\subsection{Phase 1: Supervised Pair Learning}

% alternative title: Phase 1: SSL Methods for Supervised Pair Creation

\textbf{SSL Models}\quad The aim of Phase 1 is to explore the potential ceiling of popular self-supervised learning methods in the yet-unexplored context of malware clustering (and in the tabular domain). Thus, the SSL methods used in this experiment are BYOL, SimSiam, BarlowTwins, and VICReg. All methods learn their representations by training on positive samples (samples that should be close in the representation space). The motivation for using these methods is that they should learn good representations as long as they receive strong signals from the positive pairs. At the time of writing this paper, we are not aware of any implementations of these methods for tabular data, so we created our own implementations while drawing inspiration from several GitHub repository implementations\footnote{https://github.com/PatrickHua/SimSiam}\footnote{https://github.com/facebookresearch/barlowtwins}\footnote{https://github.com/facebookresearch/vicreg}. In addition to the SSL methods, we also train an MLP classifier intending to have another \say{ceiling baseline} which can be compared with the ceilings of the SSL methods. It is trained on the task of predicting malware families (with benign samples included), which could be viewed as similar to the supervised pair-learning task using SSL methods. Afterwards, we extract its latent feature vectors and cluster them, the same as with other SSL methods.

Most of the implementation work on SSL methods consisted of replacing the (often CNN-based) model backbones with simple fully connected (FC) neural network layers, followed by a non-linear activation function. All methods use a Siamese neural network architecture. 
BYOL and SimSiam share the same loss function, whereas BarlowTwins and VICReg each have their own. The loss functions of BarlowTwins and VICReg additionally depend on one and three hyperparameters, respectively. We determine these hyperparameters based on preliminary experimentation. For BarlowTwins, we set $\lambda$ to \num{0.16} for both datasets. For VICReg, we set the hyperparameter triple $(invariance, variance, covariance)$ to $(19, 30, 1)$ for Bodmas and $(5, 100, 5)$ for Ember.

The final embeddings, which can be used for downstream tasks (e.g., clustering), can then be extracted from the Backbone or the Projector. For our case, we extract the embeddings from the Projector. Nevertheless, we set the embedding size to \num{10} for both the Backbone and the Projector. We train all of the SSL methods with a batch size of \num{1024} for \num{800} epochs and save the model weights every 50 epochs. Moreover, to ensure robust metrics, we also run each $(representation,dataset)$ combination five times.

\textbf{Supervised Pair Sampling}\quad We adopt the same supervised pair sampling approach for training all of the selected SSL methods. Let $\mathcal{D} = \{(x_i, y_i)\}^{N}_{i=1}$ represent a dataset consisting of \(N\) labeled samples, where each sample \(x_i \in \mathcal{X}\) is associated with a label \(y_i \in \mathcal{Y}\).  For the label space \(\mathcal{Y}\) it holds that \( y_i \in \{\text{malware family labels, benign label} \}\). Furthermore, we define a binary relation \( \mathcal{R} \) on the set \( \mathcal{D} \) as follows:

\[
\mathcal{R}((x_i,y_i), (x_j,y_j)) =
\begin{cases}
1, & \text{if } y_i = y_j \\
0, & \text{otherwise}
\end{cases}
\]

% The set of positive pairs \( \mathcal{P}_{+} \) is then defined as:
% %\[ 
% $ 
% \mathcal{P}_{+} = \{(x_i, x_j) \mid \mathcal{R}((x_i,y_i), (x_j,y_j)) = 1 \}
% $ 
% %\]

We define one epoch, $E$, of training the SSL methods (on positive pairs) as:

% \[
% E = \left\{ (x_i, x_j) \mid \exists j \text{ such that } \mathcal{R}(x_i, x_j) = 1, \, 

% \forall x_i \in \mathcal{D}, 1 \leq i \leq N \right\}
% \]

% \begin{align*}
% E = \left\{ (x_i, x_j) \mid \exists j \text{ such that } \mathcal{R}(x_i, x_j) = 1, \right. \\
% & \left. \forall x_i \in \mathcal{D}, 1 \leq i \leq N \right\}
% \end{align*}

$
E = \Big\{ (x_i, x_j) \mid  \exists j \text{ such that } \mathcal{R}((x_i,y_i), (x_j,y_j)) = 1,  \forall x_i \in \mathcal{D}, 1 \leq i \leq N \Big\}
$.

\subsection{Phase 2: Unsupervised Pair Learning}
\textbf{SSL \& TRL Models}\quad
Phase 2 aims to explore the potential of improving malware clustering quality in a purely unsupervised manner. This makes the SSL methods utilized in Phase 1 potentially unfit for the task, as-is, because they rely solely on learning from positive pairs. Data augmentations for tabular data, in general, do not guarantee the creation of positive (or negative) samples. In fact, in the case of tabular representation learning, the \say{data augmentations} that are employed are often referred to as data corruptions \cite{yoon2020vime,wu2024switchtab,bahri2021scarf}. Consequently, one of the tasks the TRL methods often solve is to reconstruct the feature vector of sample $x_i$ from a corrupted input $\bar{x}_i$ \cite{yoon2020vime,wu2024switchtab}.

For Phase 2, we primarily use tabular representation learning methods that do not rely on positive pairs. We utilize popular tabular representation learning methods - VIME, SCARF, and SwitchTab. Additionally, to preserve some level of comparability with Phase 1, we utilize the two best SSL methods used in Phase 1 -  BYOL and SimSiam. We train the methods for \num{800} epochs and save the model weights every \num{50} epochs, as in Phase 1. We also run each $(representation,dataset)$ combination five times to ensure robust metrics.

%For BYOL and SimSiam, the conditions are kept the same as in Phase 1, except for the sample pairs which the models train on. For 

\textbf{Unsupervised Pair Generation}\quad 
This is the most crucial step for SSL and TRL methods as it greatly influences the quality of the learned embeddings. The way unsupervised pairs are generated differs across the TRL methods we use. VIME corrupts the input features by sampling from each feature's empirical marginal distribution. Formally, we define the overall generating process of corruptions for the TRL methods as \cite{yoon2020vime}:
\begin{equation}
\label{eq:trl_corruption}
    \mathbf{\tilde{x}} = g(\mathbf{x}, \mathbf{m}) = \mathbf{m} \odot \mathbf{\bar{x}} + (1 - \mathbf{m}) \odot \mathbf{x}
\end{equation} % _{\text{VIME}}
For all of the TRL methods, $m$ corresponds to a binary masked vector $ m = \left[ m_1, \ldots, m_d \right] $ sampled from a Bernoulli distribution with a probability of $p_m$. In the case of VIME, the $j$-th feature of $\bar{x}$ is sampled from the empirical distribution 
% $\hat{p}_{X_j} = \frac{1}{N_u} \sum_{i=N_l+1}^{N_l+N_u} \delta(x_j = x_{i,j}) $
$\hat{p}_{X_j} = \frac{1}{N} \sum_{i=1}^{N} \delta(x_j = x_{i,j})$
where $x_{i,j}$ is the $j$-th feature of the $i$-th sample in $\mathcal{D}$ (i.e., the empirical marginal distribution of each feature). We set VIME's rate of corruption, $p_m$, to $0.3$, the same as in the original work \cite{yoon2020vime}. The same pair creation/corruption approach is used to generate pairs for BYOL and SimSiam.

The same general description of a corruption formula applies to SCARF as in Equation \ref{eq:trl_corruption}. Moreover, SCARF originally uses the same approach for data corruption as VIME. However, during preliminary experimentation, we discovered that it learns better with a different corruption approach. Therefore, for SCARF, the $j$-th feature of $\bar{x}$ is sampled from a uniform distribution created from the minimum and maximum of the feature values, specifically $\bar{x_j} \sim \mathcal{U}\left(\min(x_j), \max(x_j)\right)$. We use a corruption rate, $ p_m$, of \num{0.3} for SCARF.
% as specified by the original authors \cite{bahri2021scarf}.
% We used the same corruption rate, $ p_m = 0.6$, for SCARF as specified by the original authors \cite{bahri2021scarf}.

SwitchTab, on the other hand, uses a corruption-by-zero approach, in principle the same as for Denoising Autoencoders. Therefore, we can also define its corruption process using Equation \ref{eq:trl_corruption} where the whole corrupted feature vector is a vector of zeros, i.e. $\bar{\mathbf{x}} = \mathbf{0}_d$, where $d$ is the dimensionality of the feature space. The feature corruption ratio, $p_m$, was set to $0.3$ as in the original work \cite{wu2024switchtab}.

\subsection{Binary Program Clustering}
After the representations from Phases 1 and 2 have been trained, we extract 10-dimensional embeddings and cluster them using a clustering algorithm. Inspired by the positive results using K-Means in \cite{mocko2025clustering}, we employ the K-Means algorithm via the CUDA-accelerated \texttt{cuML} library. 
Unlike the study of \cite{mocko2025clustering}, however, we do not utilize an amount of clusters that would be much higher than the reported number of malware families. Instead, we keep the number of clusters on the level of the number of malware families in the train portion of the respective datasets - 560 in the case of Bodmas and 2750 in the case of Ember.
% Following the experimental configuration specified in the reference work, the cluster count $k$ is set to \num{2000} for the Bodmas dataset and \num{50000} for the Ember dataset. Maintaining these specific parameters provides a consistent evaluation framework across the two studies (as no other study has been conducted on the full set of datasets used). 
The remaining hyperparameters are kept at their default values, specifically 300 iterations ($n\_iter$) and the scalable k-means++ initialization scheme. The k-means++ initialization scheme improves convergence stability and reduces the variance of the final clustering error.
We cluster all representations saved after every 50 epochs of SSL/TRL training for each run of the $(representation,dataset)$ combinations. Afterwards, we perform clustering quality evaluation as described in Section \ref{sec:evaluation_metrics}. 

%Based on the results presented in \cite{mocko2025clustering} as well as reasonable time complexity, we choose the K-Means algorithm to cluster the malware datasets. More specifically, we choose the \texttt{cuML}\footnote{https://github.com/rapidsai/cuml} implementation that utilizes CUDA. Following the best reported results in \cite{mocko2025clustering}, we set $k$, the number of clusters, to  $\num{2000}$ for Bodmas and $\num{50000}$ for Ember. We leave the other hyperparameters unchanged, leaving $n\_iter$ set to \num{300}, and scalable kmeans++ initialization. 

\subsection{Evaluation \& Setup}
\label{sec:evaluation_metrics}
We view the utility and quality of a malware clustering model primarily in its function to group samples that are strongly correlated. 
% Moreover, we recognize the heterogeneity in PE samples. Thus, \emph{we do not deem it necessary to group a single malware family into a single cluster}, as there may be legitimate reasons why samples within the family differ \cite{hu2013duet} (e.g., sub-variants, different behaviors, different threat actors using them for different purposes, etc.). 
Therefore, we determine our primary metric to be Homogeneity. As stated in Section \ref{sec:introduction}, in the literature, there are plenty of other malware clustering works that place similar significance on the metric of Homogeneity/Purity \cite{fang2019semi,wilkins2020cougar,jurevckova2024classification,jurevckova2025online,mocko2025clustering}. Given that $H$ represents the Shannon Entropy, we define Homogeneity $h$ as:
\begin{equation}
\label{eq:metrics_homogeneity}
    h = 
    \begin{cases}
        1, & \text{if } H(C,K) = 0 \\
        1 - \frac{H(C|K)}{H(C)}, & \text{otherwise}
    \end{cases}
\end{equation}

where $H(C)$ represents class entropy of the true class labels and $H(C|K)$ represents Conditional Entropy of the true class labels, $C$, given the cluster assignments, $K$.

To also ensure a more robust evaluation, as well as to provide readers with additional metric information, we also compute Completeness and V-Measure. 
% To achieve a thorough evaluation, we compute Homogeneity 
All of the reported metrics are computed for both the \emph{class labels} (malware/benign) and the \emph{family labels}. We define \say{family labels} for \emph{all} samples in the dataset. Malware samples are assigned fine-grained family labels (e.g., Emotet, TrickBot), while all legitimate software is assigned the single, coarse-grained label of \say{benign}. 

As mentioned in the previous subsection, clustering is computed every 50 epochs for the methods used (except for the baselines, for some of which this is not possible). Using 800 epochs for training yields 16 measurements (checkpoints) for each method. Moreover, there are five runs for each $(representation,dataset)$ combination. This effectively results in 80 total model checkpoints. 
To ensure objective clustering quality evaluation, we report robust metrics for both Phase 1 and Phase 2. For Phase 1, we report the maximum Homogeneity (other metrics as well) per run, averaged across five runs, together with standard deviation. 
This is natural since in Phase 1 we operate in a setting where label information is available. Therefore, it is easy to pick the best-trained model state (for each run of the representation methods). 

In Phase 2, the situation is more complicated since we don't assume access to labeled information where a practitioner could pick the best-trained model state based on such information. Therefore, we report mean Homogeneity averaged across five model runs (each run consists of 16 checkpoints), together with its standard deviation. Moreover, we also report maximum Homogeneity in the fashion as for Phase 1, also together with the standard deviation. These metrics are computed on the train set of the datasets. Other metrics like Completeness and V-Measure are also reported in the same fashion.

We run our experiments on a machine with 26 CPU cores, 225 GB of RAM, 1 Nvidia H100 GPU, 1 TB of disk space, and a Linux LTS 24.04 operating system. We divide the H100 GPU into 4 MiG instances to effectively isolate and run experiments in parallel across 4 separate GPU parts.

% \todo[inline]{Important Hyperparams? Learning rate, optimizer, batch size, }

% besides other stuff, we should also provide infomrmation on how we set up the methods, if they needed some preliminary experiments
% basically, for BT we should show that we needed to change the coef to 0.16 and that there is also some kind of relationship there which was helpful
% I think it was something like their dataset size / our dataset size * something .. or maybe it was something else
% for VICReg it was a lot of blood sweat and tears

% defend the use of single clustering method - K-Means
% our goal in this work is not to exhaustively search for the best clustering algo but to pick one that is decent and show how the results change when changing the input representation

% also we want to show that the representation is the key to the clustering success and that the results may very in the tens of percents due to maybe an unfit representation

% VICReg paper claims that BYOL and Barlow Twins both utilize LARS optimizer; VICReg uses it as well
% SimSIam - they use SGD optimizer !! 

\section{Experimental Results}
\label{sec:experiments}

In this section, we present an evaluation of the conducted experiments. We discuss and give answers to each of our research questions: (RQ1) Supervised Pair Learning: SSL model comparison for malware clustering compared with unsupervised state-of-the-art baselines, and
(RQ2) Unsupervised Pair Learning: comparison of state-of-the-art unsupervised malware clustering baselines together with new unsupervised TRL approaches.

\subsection{Phase 1: Supervised Pair Learning}
\label{sec:experiments_phase1}

% \begin{table}[!h]
% \centering
% \caption{Results for Phase 1 - Supervised Pair Learning. Performance of baselines and SSL methods reported via Homogeneity on class and family labels.}
% \label{tab:phase1_results}
% \footnotesize
% \begin{tabular}{l @{\hspace{1em}} lcc}
% \toprule
% \textbf{Dataset} & \textbf{Method} & \textbf{Homog. Class $\uparrow$} & \textbf{Homog. Family $\uparrow$} \\
% \midrule
% \multirow{9}{*}{Bodmas} 
%  & PCA           & 89.47\% & 82.00\% \\
%  & UMAP          & 92.56\% & \underline{89.88\%} \\
%  & AE 300        & 89.61\% & 83.91\% \\
%  & AE 800        & 90.46\% & 84.39\% \\
%  & BarlowTwins   & 89.09\% & 82.85\% \\
%  & BYOL          & \underline{98.51\%} & 88.49\% \\
%  & MLPClassifier & \textbf{99.12\%} & \textbf{92.54\%} \\
%  & SimSiam       & 97.24\% & 87.91\% \\
%  & VICReg        & 94.46\% & 70.00\% \\
% \midrule
% \multirow{9}{*}{Ember} 
%  & PCA           & 81.60\% & 87.68\% \\
%  & UMAP          & 79.77\% & 86.74\% \\
%  & AE 300        & 81.44\% & 87.69\% \\
%  & AE 800        & 81.30\% & 87.63\% \\
%  & BarlowTwins   & 80.52\% & 87.74\% \\
%  & BYOL          & \textbf{94.90\%} & \underline{90.00\%} \\
%  & MLPClassifier & 94.14\% & 86.56\% \\
%  & SimSiam       & \underline{94.62\%} & \textbf{90.30\%} \\
%  & VICReg        & 55.52\% & 62.34\% \\
% \bottomrule
% \end{tabular}
% \end{table}

\begin{table}[t]
\centering
\caption{Results for Phase 1 - Supervised Pair Learning. Performance of baselines and SSL methods reported via Homogeneity on class and family labels. Values are reported in percentages.}
% \caption{Phase 1: Homogeneity analysis of learned representations on the training set. \textbf{Bold} indicates the best result per dataset; \underline{underline} indicates the second best.}
\label{tab:phase1_results}
% \resizebox{0.75\textwidth}{!}{%
\begin{tabular}{ll cc}
\toprule
& & \multicolumn{2}{c}{\textbf{Homogeneity (Mean of Maxes)}} \\
\cmidrule(lr){3-4}
\textbf{Dataset} & \textbf{Method} & Class-level & Family-level \\
\midrule
\multirow{8}{*}{BODMAS}
& PCA          & 80.25 $\pm$ 1.09 & 71.96 $\pm$ 0.79 \\
& Autoencoder  & 84.48 $\pm$ 0.32 & 76.51 $\pm$ 0.30 \\
& UMAP         & 82.34 $\pm$ 0.60 & 80.43 $\pm$ 0.28 \\
& Barlow Twins & 77.40 $\pm$ 15.17 & 68.39 $\pm$ 13.43 \\
& BYOL         & \textbf{99.42} $\pm$ 0.32 & \underline{84.92} $\pm$ 1.61 \\
& MLP          & 97.94 $\pm$ 0.53 & \textbf{87.85} $\pm$ 1.18 \\
& SimSiam      & \underline{98.46} $\pm$ 1.05 & 80.86 $\pm$ 2.01 \\
& VICReg       & 38.88 $\pm$ 53.23 & 28.85 $\pm$ 39.55 \\
\midrule
\multirow{7}{*}{EMBER}
& PCA          & 58.02 $\pm$ 0.19 & 66.13 $\pm$ 0.24 \\
& Autoencoder  & 58.80 $\pm$ 0.12 & 67.44 $\pm$ 0.23 \\
& UMAP         & 67.13 $\pm$ 0.32 & 75.55 $\pm$ 0.12 \\
& Barlow Twins & 60.97 $\pm$ 2.96 & 61.41 $\pm$ 5.21 \\
& BYOL         & \underline{91.36} $\pm$ 2.64 & \underline{75.87} $\pm$ 0.60 \\
& MLP          & \textbf{96.09} $\pm$ 0.54 & 66.01 $\pm$ 6.86 \\
& SimSiam      & 90.40 $\pm$ 2.28 & \textbf{76.53} $\pm$ 2.20 \\
& VICReg       & 42.54 $\pm$ 16.23 & 47.63 $\pm$ 12.10 \\
\bottomrule
\end{tabular}%
% }
\end{table}

Table \ref{tab:phase1_results} summarizes the results for Phase 1 of our experiments. We present only the best results, as this task assumes the availability of labels; thus, choosing the best-trained representation is straightforward (based on label information; as opposed to unsupervised representation learning and clustering, where we presume no label availability). The best results are represented in the form of a \say{mean of maxes}, i.e. maximum Homogeneity that is averaged across five runs for all of the models. We include all the unsupervised baselines from \cite{mocko2025clustering}. The Autoencoder baseline is now, however, trained on 800 epochs to align the results with SSL models, which all utilized 800 epochs for training. Moreover, we train an MLP classifier (for 800 epochs) to evaluate results from a fully supervised perspective, directly optimizing for the malware family classification task. After the MLP classifier is trained, its internal latent representation is extracted and used as input to the clustering algorithm.

% \textbf{MLP does not always win.}\quad Even though the latent feature representation from the MLP classifier may be considered the clear favorite because the MLP is trained to directly optimize for classifying all the malware families (and benign samples), it surprisingly does not always achieve the best results. In the case of Bodmas, it clearly beats all the competition with 99.12\% Homogeneity on class and 92.54\% Homogeneity on family. However, on the Ember dataset, the MLP struggles, ranking only third for Homogeneity on class (94.14\%) and second-to-last for Homogeneity on family. 

\textbf{MLP does not always win.}\quad Despite being optimized directly for malware classification, the MLP-derived latent representations do not consistently yield superior clustering results in all aspects. On Bodmas, MLP dominates family Homogeneity with 87.85\%, an almost 3\% lead over the second best performer, BYOL. However, MLP's performance is only top-3 in class Homogeneity on said dataset. Even though MLP can dominate class Homogeneity on Ember, its performance diminishes on family Homogeneity, arguably the more important indicator of overall clustering quality. These findings suggest that MLP's discriminative power does not always generalize across different malware distributions.

\textbf{BYOL outperforms other SSL methods.}\quad Among the four SSL methods utilized in Phase 1 experiments, BYOL is the most consistent in achieving top results. It achieves the best results in terms of Homogeneity on class (99.42\%) and Homogeneity on family (84.92\%) among the four SSL methods on Bodmas. On Ember, it even outperforms MLP on family Homogeneity by a very large margin, more than 9\%. In addition, SimSiam deserves an honorable mention, ranking second among the four SSL methods. SimSiam is even able to beat other models in family Homogeneity on Ember. Overall BYOL and SimSiam seem to perform relatively similarly. 
% On the other hand, it is worth noting that none of the SSL methods can beat the \emph{unsupervised} UMAP baseline on family Homogeneity for Bodmas, which we deem quite surprising. On the other hand, this is not the case on the Ember dataset.

% \textbf{BarlowTwins and VICReg underperform.}\quad These two methods differ from BYOL and SimSiam in their loss functions and, based on our experiments, seem to significantly underperform their aforementioned counterparts. The underperformance can be seen on Homogeneity on class and on family as well. It is best illustrated on Homogeneity on family, where, on Bodmas, BarlowTwins cannot even achieve the performance of an Autoencoder trained for 300 epochs (82.85\% vs 83.91\%). Therefore it is not even able to beat an \emph{unsupervised} baseline, let alone a supervised approach. The results for VICReg are even more pronounced, where the Homogeneity on family is the worst overall on both datasets, with 70.00\% on Bodmas and 62.34\% on Ember. This behavior of the two models may be caused by not-optimally-chosen coefficients for their loss functions, however, we spent significant effort in preliminary experiments on trying to find at least reasonable coefficients where the methods learn reasonably and not reach model collapse. 

\textbf{Barlow Twins and VICReg Underperform.}\quad Unlike BYOL and SimSiam, these methods significantly underperformed across all metrics. Both of these methods did not manage to beat the baselines on Bodmas. It is only on Ember that BarlowTwins is able to beat PCA and Autoencoder in class Homogeneity. VICReg performed worst overall, with its performance not being able to keep up with even the weakest of baselines. The high standard deviations say it all - the methods are hit-and-miss. Despite extensive tuning to prevent model collapse or poor performance, we were unable to make BarlowTwins and VICReg perform consistently well across multiple runs. If this is not possible to do while providing them with the best signal they can get (supervised pairs), it is not a worthwhile effort to try and make them work in an unsupervised context (Phase 2).

\textbf{Supervised Pair Learning ceiling for binary program clustering established.}\quad Looking at the results of the experiments from a high-level perspective, three major top models seem to emerge - MLP classifier, BYOL, and SimSiam, although not without some occasional hiccups (mainly MLP underperforming on family Homogeneity for Ember). Based on the results, we establish the current ceiling for supervised pair learning (or fully supervised learning) for both datasets. On Bodmas, it is around 99\% class Homogeneity and 87.85\% family Homogeneity. On Ember, it is just over 96\% class Homogeneity and 76.53\% family Homogeneity. This, of course, holds true for the respective amounts of clusters that we utilized (560 for Bodmas, 2750 for Ember). The established ceilings serve as useful guides for assessing the remaining headroom for the unsupervised approaches used in Phase 2, as well as for any future experiments where researchers might want to try out their own custom data augmentation techniques.

\begin{tcolorbox}[
    boxsep=3pt,          % Space between text and box border
    left=5pt,            % Left padding
    right=5pt,           % Right padding
    top=5pt,             % Top padding
    bottom=5pt,          % Bottom padding
    colback=white!15!gray!7, % Light grey background
    colframe=black!125,   % Slightly darker grey frame
    arc=1mm,             % Rounded corners
    outer arc=1mm,       % Outer rounded corners
    boxrule=0.4pt,       % Thickness of the frame
    fonttitle=\bfseries, % Make title bold (if you use a title)
    % title={\textbf{Malware Classification Accuracy:}} % Example title
]
\textbf{RQ1 Finding:}
The four utilized SSL methods are not created equal. BYOL and SimSiam greatly outperform BarlowTwins and VICReg on the task of binary program clustering when trained based on supervised pairs.
\end{tcolorbox}

% \usepackage{multirow}
% \usepackage{graphicx}
% \resizebox{\textwidth}{!}

% \begin{tcolorbox}[
%     boxsep=3pt,          % Space between text and box border
%     left=5pt,            % Left padding
%     right=5pt,           % Right padding
%     top=5pt,             % Top padding
%     bottom=5pt,          % Bottom padding
%     colback=white!15!gray!7, % Light grey background
%     colframe=black!125,   % Slightly darker grey frame
%     arc=1mm,             % Rounded corners
%     outer arc=1mm,       % Outer rounded corners
%     boxrule=0.4pt,       % Thickness of the frame
%     fonttitle=\bfseries, % Make title bold (if you use a title)
%     % title={\textbf{Malware Classification Accuracy:}} % Example title
% ]
% \textbf{Finding 1:}
% Method X was able to outperform existing state-of-the-art deep learning models for malware-type and malware-family with an macro F1-Score of .497 and .491 respectively.
% \end{tcolorbox}

\subsection{Phase 2: Unsupervised Pair Learning}

\begin{table*}[t]
\centering
\caption{Phase 2 Results: Homogeneity and V-Measure analysis of learned unsupervised representations. We compare the average (Mean) and best (Max) performance across SSL and TRL methods on the Bodmas and Ember datasets, averaged over five experiment runs. Values are reported in percentages.} % \textbf{Bold} indicates the best result per dataset; \underline{underline} indicates the second best.
\label{tab:phase2_combined}
\resizebox{\textwidth}{!}{%
\begin{tabular}{ll cccc cccc}
\toprule
& & \multicolumn{4}{c}{\textbf{Homogeneity} $\uparrow$} & \multicolumn{4}{c}{\textbf{V-Measure} $\uparrow$} \\
\cmidrule(lr){3-6} \cmidrule(lr){7-10}
& & \multicolumn{2}{c}{Class-level} & \multicolumn{2}{c}{Family-level} & \multicolumn{2}{c}{Class-level} & \multicolumn{2}{c}{Family-level} \\
\cmidrule(lr){3-4} \cmidrule(lr){5-6} \cmidrule(lr){7-8} \cmidrule(lr){9-10}
\textbf{Dataset} & \textbf{Method} & Mean of Means & Mean of Maxes & Mean of Means & Mean of Maxes & Mean of Means & Mean of Maxes & Mean of Means & Mean of Maxes \\
\midrule
\multirow{8}{*}{BODMAS}
& PCA          & 80.25 $\pm$ 1.09 & 80.25 $\pm$ 1.09 & 71.96 $\pm$ 0.79 & 71.96 $\pm$ 0.79 & 18.14 $\pm$ 0.21 & 18.14 $\pm$ 0.21 & 44.28 $\pm$ 0.40 & 44.28 $\pm$ 0.40 \\
& Autoencoder  & 82.37 $\pm$ 0.31 & 84.48 $\pm$ 0.32 & 74.60 $\pm$ 0.32 & 76.51 $\pm$ 0.30 & 18.24 $\pm$ 0.08 & 18.67 $\pm$ 0.07 & 45.17 $\pm$ 0.22 & 46.06 $\pm$ 0.18 \\
& UMAP         & 82.34 $\pm$ 0.60 & 82.34 $\pm$ 0.60 & \underline{80.43} $\pm$ 0.28 & \underline{80.43} $\pm$ 0.28 & 17.73 $\pm$ 0.16 & 17.73 $\pm$ 0.16 & \underline{47.63} $\pm$ 0.25 & \underline{47.63} $\pm$ 0.25 \\
& SCARF        & \underline{84.35} $\pm$ 0.41 & \underline{85.50} $\pm$ 0.28 & 76.41 $\pm$ 0.15 & 76.93 $\pm$ 0.15 & \underline{18.51} $\pm$ 0.08 & \underline{18.77} $\pm$ 0.06 & 45.94 $\pm$ 0.08 & 46.24 $\pm$ 0.06 \\
& SwitchTab    & 82.93 $\pm$ 0.45 & 84.17 $\pm$ 0.41 & 79.10 $\pm$ 0.42 & 79.78 $\pm$ 0.47 & 17.73 $\pm$ 0.05 & 18.04 $\pm$ 0.07 & 46.59 $\pm$ 0.14 & 47.17 $\pm$ 0.19 \\
& VIME         & \textbf{87.02} $\pm$ 0.62 & \textbf{87.85} $\pm$ 0.66 & \textbf{80.87} $\pm$ 0.22 & \textbf{81.66} $\pm$ 0.37 & \textbf{18.94} $\pm$ 0.23 & \textbf{19.13} $\pm$ 0.26 & \textbf{48.30} $\pm$ 0.33 & \textbf{48.78} $\pm$ 0.43 \\
& $\text{BYOL}_\text{VIME}$    & 67.52 $\pm$ 5.02 & 73.11 $\pm$ 3.91 & 68.22 $\pm$ 3.21 & 71.71 $\pm$ 2.53 & 15.35 $\pm$ 1.10 & 16.46 $\pm$ 1.03 & 42.14 $\pm$ 1.49 & 43.90 $\pm$ 1.35 \\
& $\text{SimSiam}_\text{VIME}$ & 72.60 $\pm$ 2.00 & 77.80 $\pm$ 1.61 & 72.73 $\pm$ 1.44 & 75.25 $\pm$ 1.11 & 16.03 $\pm$ 0.40 & 17.11 $\pm$ 0.36 & 43.93 $\pm$ 0.74 & 45.25 $\pm$ 0.61 \\
\midrule
\multirow{8}{*}{EMBER}
& PCA          & 58.02 $\pm$ 0.19 & 58.02 $\pm$ 0.19 & 66.13 $\pm$ 0.24 & 66.13 $\pm$ 0.24 & 11.00 $\pm$ 0.02 & 11.00 $\pm$ 0.02 & 41.96 $\pm$ 0.10 & 41.96 $\pm$ 0.10 \\
& Autoencoder  & 58.32 $\pm$ 0.30 & 58.80 $\pm$ 0.12 & 67.08 $\pm$ 0.37 & 67.44 $\pm$ 0.23 & 10.97 $\pm$ 0.04 & 11.05 $\pm$ 0.03 & 42.31 $\pm$ 0.17 & 42.49 $\pm$ 0.14 \\
& UMAP         & \textbf{67.13} $\pm$ 0.32 & \textbf{67.13} $\pm$ 0.32 & \textbf{75.55} $\pm$ 0.12 & \textbf{75.55} $\pm$ 0.12 & 11.36 $\pm$ 0.05 & 11.36 $\pm$ 0.05 & \underline{43.94} $\pm$ 0.06 & 43.94 $\pm$ 0.06 \\
& SCARF        & 60.41 $\pm$ 0.05 & 60.82 $\pm$ 0.06 & 69.24 $\pm$ 0.05 & 69.45 $\pm$ 0.05 & 11.33 $\pm$ 0.01 & 11.40 $\pm$ 0.01 & 43.57 $\pm$ 0.02 & 43.72 $\pm$ 0.04 \\
& SwitchTab    & 64.56 $\pm$ 0.59 & 65.29 $\pm$ 0.67 & 72.03 $\pm$ 0.19 & 72.57 $\pm$ 0.24 & \underline{11.57} $\pm$ 0.07 & \underline{11.70} $\pm$ 0.08 & 43.79 $\pm$ 0.14 & \underline{44.06} $\pm$ 0.14 \\
& VIME         & \underline{66.04} $\pm$ 0.41 & \underline{66.80} $\pm$ 0.38 & \underline{73.61} $\pm$ 0.50 & \underline{74.36} $\pm$ 0.32 & \textbf{12.12} $\pm$ 0.11 & \textbf{12.23} $\pm$ 0.11 & \textbf{45.59} $\pm$ 0.16 & \textbf{45.81} $\pm$ 0.18 \\
& $\text{BYOL}_\text{VIME}$    & 51.09 $\pm$ 1.61 & 52.92 $\pm$ 1.63 & 64.82 $\pm$ 1.10 & 66.24 $\pm$ 1.28 & 9.41 $\pm$ 0.20 & 9.78 $\pm$ 0.27 & 40.24 $\pm$ 0.40 & 41.02 $\pm$ 0.45 \\
& $\text{SimSiam}_\text{VIME}$ & 53.55 $\pm$ 0.74 & 56.41 $\pm$ 0.92 & 67.34 $\pm$ 0.43 & 68.86 $\pm$ 0.87 & 9.79 $\pm$ 0.12 & 10.17 $\pm$ 0.12 & 41.57 $\pm$ 0.23 & 42.16 $\pm$ 0.19 \\
\bottomrule
\end{tabular}%
}
\end{table*}

This section covers experiments conducted for Phase 2 using SCARF, SwitchTab, and VIME as promising tabular representation learning methods. Table \ref{tab:phase2_combined} shows both the mean and the maximum Homogeneity of the Phase 2 methods (together with unsupervised baselines) averaged over five method runs (i.e. \say{Mean of Means} and \say{Mean of Maxes}). This is also the case for V-Measure. We remind the reader that one run results in 16 model snapshots (i.e. checkpoints) collected during the training over 800 epochs. The main reason for showing the mean of the Homogeneities (or other metrics) is that it better captures the average performance of the models when no reasonable guess can be made about the point (i.e., the epoch) at which the model training should be stopped (and the model selected). 
% This can be very informative for the scenario when no labels are assumed to be available for model selection. 
Difficulty in identifying the optimal training stopping point is common during pure unsupervised training, especially when the model loss is poorly correlated with downstream task performance. The maximum Homogeneity showcases the representation's full capabilities. 
% \todo[inline]{Perhaps add a comment that more computed metrics can be found in the attachments (if we want to add them)}

Moreover, we also use the two best-performing SSL models from Phase 1 (Section \ref{sec:experiments_phase1}), i.e., BYOL and SimSiam, for these experiments. Since this is an unsupervised pair learning experiment, both BYOL and SimSiam receive pairs via VIME's pair creation mechanism, i.e., corruption via sampling from the empirical marginal distribution. This showcases the mechanism purely as a \say{data augmentation} technique aimed at creating positive samples, since BYOL and SimSiam do not employ any reconstruction or mask-prediction mechanism, as in VIME.

\textbf{VIME beats strong unsupervised baselines.}\quad VIME achieves state-of-the-art results in Phase 2 experiments. Moreover, VIME consistently performs best among the tested TRL methods. 
On Bodmas, it beat out all unsupervised baselines and all TRL method contenders in a strong fashion - with a 2.35\% lead in class Homogeneity and 1.2\% lead in family Homogeneity. On Ember, it manages to get very close to the surprisingly strong UMAP baseline, achieving a class Homogeneity of 66.80\% and family Homogeneity of 74.36\%. UMAP was surprisingly strong on Bodmas as well, showing that it is, in fact, a very strong baseline and that UMAP belongs to the repertoire of modern representation learning methods. These very strong top-1 and top-2 placements of VIME point to the data corruption technique used by VIME as very effective for unsupervised tabular representation learning, and to VIME itself as a state-of-the-art method for binary program clustering. 

\textbf{SwitchTab claims second place among TRL methods.}\quad SwitchTab is a very strong contender to VIME primarily in family Homogeneity, where their results are often very similar (around 1\% to 2\% apart). Even though SwitchTab never beat VIME, it is very close in performance mainly on the Ember dataset. The performance difference between VIME and SwitchTab is more significant on class Homogeneity, where SwitchTab performs 1.5\%-4\% worse. Overall, SwitchTab deserves its place as the second-best and shows that the use of corruption by zeros, paired with decoupling mutual and salient features, helps learn highly informative unsupervised representations.

% \vspace{-3pt}
\begin{tcolorbox}[
    boxsep=3pt,          % Space between text and box border
    left=5pt,            % Left padding
    right=5pt,           % Right padding
    top=5pt,             % Top padding
    bottom=5pt,          % Bottom padding
    colback=white!15!gray!7, % Light grey background
    colframe=black!125,   % Slightly darker grey frame
    arc=1mm,             % Rounded corners
    outer arc=1mm,       % Outer rounded corners
    boxrule=0.4pt,       % Thickness of the frame
    fonttitle=\bfseries, % Make title bold (if you use a title)
    % title={\textbf{Malware Classification Accuracy:}} % Example title
]
\textbf{RQ2 Finding 1:}
Tabular representation learning methods can beat strong unsupervised baselines and improve the purely unsupervised quality of deep representations for binary program clustering, achieving new state-of-the-art results.
\end{tcolorbox}

% \textbf{VIME versus Supervised Pair Results.}\quad One of the most interesting observations is that VIME beats the MLP classifier on Ember in family Homogeneity. Moreover, the difference between the best achieved family Homogeneity on Ember using supervised pair learning (SimSiam, Table \ref{tab:phase1_results}) and VIME is only around 2\% (considering the maximum results in Table \ref{tab:phase2_combined}), which does not constitute a large difference because SimSiam received all the important signals about which pairs should be close together, which VIME did not. 
% Another interesting observation is that VIME was only slightly worse than BYOL and SimSiam on family Homogeneity on Bodmas. 
% Moreover, in the case of Bodmas, SwitchTab even beat SimSiam on family Homogeneity by 0.29\%. This, in our opinion, 1) proves that SwitchTab and VIME are very good unsupervised learners and 2) points us to speculate that the representation learning capability of BYOL and SimSiam on tabular data (when trained on supervised pairs) could be improved. However, the mechanisms for achieving this improvement require further investigation and are beyond the scope of this work. 

\textbf{VIME versus Supervised Pair Results.}\quad A striking observation is that VIME, despite being fully unsupervised, comes remarkably close to the supervised ceiling on family Homogeneity. On Bodmas, the gap between VIME (81.66\% max) and the best supervised method (MLP, 87.85\%) is only 6.2\%. On Ember, this gap narrows further: VIME achieves 74.36\% (max) versus SimSiam's ceiling of 76.53\%, a difference of merely 2.2\%. This is particularly noteworthy given that SimSiam received explicit family-level supervision for pair generation, while VIME operated without any label information. These results demonstrate that VIME's corruption-based learning captures much of the discriminative structure that supervised methods exploit, and suggest that improvements to the corruption strategy could close the remaining gap entirely.

% \vspace{-2pt}
\textbf{SCARF: consistent but limited performance.}\quad SCARF ranks third among the tested TRL methods, underperforming compared to SwitchTab and VIME in family Homogeneity on both Bodmas and Ember. However, it consistently outperforms the baselines of PCA and Autoencoder on both datasets and exhibits remarkable stability; with its standard deviation in mean and maximum family Homogeneity on both datasets being very low. These results suggest that while contrastive loss on unsupervised pairs may be sub-optimal for the high similarity found in binary program clusters, SCARF remains a robust, consistent choice for broader tabular domains.

% SCARF sovereignly gains its third place out of the three tested TRL methods, making it the worst of the three. However, it still managed to beat some unsupervised baselines, which was ultimately the main goal. SCARF achieves reasonable class Homogeneity on Bodmas, but its family Homogeneity is a few percentage points lower than that of SwitchTab and VIME. In the case of Ember, SCARF outperforms all unsupervised baselines on both class and family Homogeneity. Interestingly, SCARF is also very consistent - the most consistent of the three TRL methods. This can, for example, be seen in the difference between the mean and maximum family Homogeneity on Ember, which is merely 0.04\%.
% Overall, SCARF's usage of the contrastive loss function on unsupervised pairs may not be the right fit for the binary program clustering task; however, in other domains where the underlying sample classes are less similar, it may still be a promising choice. 

% \vspace{-2pt}
\textbf{$\text{BYOL}_{\text{VIME}}$ and $\text{SimSiam}_{\text{VIME}}$ reveal both good and bad news.}\quad The bad news about BYOL's and SimSiam's performance when utilizing VIME's data augmentation for pair learning is that they are, on average, not able to reach the performance of even our unsupervised baselines. Their best performance (according to Table \ref{tab:phase2_combined}) sometimes matches that of PCA and Autoencoder baselines. However, this holds mainly for family Homogeneity and does not hold for class Homogeneity, where both $\text{SimSiam}_{\text{VIME}}$ and $\text{BYOL}_{\text{VIME}}$ often lose by more than 5 percentage points. The good news, on the other hand, is that the VIME \say{data augmentation} was strong enough for $\text{BYOL}_{\text{VIME}}$ and $\text{SimSiam}_{\text{VIME}}$ to reach a relatively good performance, considering that we presume to be learning on positive pairs; which is not necessarily the case with VIME augmentation (corruption) strategy.

% \vspace{-3pt}
\begin{tcolorbox}[
    boxsep=3pt,          % Space between text and box border
    left=5pt,            % Left padding
    right=5pt,           % Right padding
    top=5pt,             % Top padding
    bottom=5pt,          % Bottom padding
    colback=white!15!gray!7, % Light grey background
    colframe=black!125,   % Slightly darker grey frame
    arc=1mm,             % Rounded corners
    outer arc=1mm,       % Outer rounded corners
    boxrule=0.4pt,       % Thickness of the frame
    fonttitle=\bfseries, % Make title bold (if you use a title)
    % title={\textbf{Malware Classification Accuracy:}} % Example title
]
\textbf{RQ2 Finding 2:}
Positive-pair-based SSL methods using VIME data augmentation fail to match the performance of unsupervised baselines and state-of-the-art tabular representation learning methods on the binary program clustering task. 

% VIME data corruption (augmentation) strategy is powerful enough to make BYOL and SimSiam learn somewhat decent embeddings even when they expect positive-only pairs, proving that it 
\end{tcolorbox}

\subsection{VIME-R: Retrieval-Augmented Corruption of Binary Programs}
\subsubsection{Motivation}
% \textbf{} \quad 
The results from Phase 2 identify VIME as the strongest unsupervised method among those that were tested. However, VIME's corruption mechanism samples from marginal distributions without regard to sample similarity (as well as class label or malware family information). 
% \todo[inline]{ADD STUFF HERE?}
We hypothesize that retrieval-based corruption, replacing features with values from semantically similar samples, can produce more informative augmentations, even more so in the domain of malware. In theory, the samples that are the closest to another sample have the best chances of belonging to the same, or a very similar, malware family. We speculate that this could, therefore, be an optimization of VIME for improving the model's recognition of malware families. 

The hypothesis is further supported by an analysis of nearest neighbors on the Ember dataset presented in Table \ref{tab:neighbor_purity}. It clearly shows that the closest neighbors which share the same label are usually found somewhere in the Top-1 to Top-100 range. After that point, label purity drops off. 

We thus propose VIME-R, Value Imputation and Mask Estimation with Retrieval. The method is a modification of VIME where the corruption process is augmented by sampling feature values only from the marginal distribution of the k nearest neighbors of a given sample, rather than from a global marginal distribution. To our best knowledge, ours is the first work to utilize this alternative approach for the corruption process of VIME.

VIME-R modifies the corruption process by restricting the sampling pool to a sample's local neighborhood. Let $\mathcal{N}_k(x_i) = \{x_{i_1}, x_{i_2}, \ldots, x_{i_k}\}$ denote the set of $k$ nearest neighbors of sample $x_i$ in $\mathcal{D}$. The corruption process follows the same general form as Equation \ref{eq:trl_corruption}:
\begin{equation}
\label{eq:vimer_corruption}
    \mathbf{\tilde{x}}_i = g(\mathbf{x}_i, \mathbf{m}) = \mathbf{m} \odot \mathbf{\bar{x}}_i + (1 - \mathbf{m}) \odot \mathbf{x}_i
\end{equation}
However, unlike VIME, which samples from the global empirical marginal distribution, VIME-R samples the $j$-th feature of $\bar{x}_i$ from the local empirical marginal distribution conditioned on the neighborhood $\mathcal{N}_k(x_i)$:
\begin{equation}
\label{eq:vimer_local_marginal}
    \hat{p}_{X_j | \mathcal{N}_k(x_i)} = \frac{1}{k} \sum_{l=1}^{k} \delta(x_j = x_{i_l, j})
\end{equation}
where $x_{i_l, j}$ is the $j$-th feature of the $l$-th nearest neighbor of $x_i$. The mask vector $\mathbf{m}$ is sampled identically to VIME from a Bernoulli distribution with probability $p_m$.

The aim for VIME-R is to be a purely unsupervised method in our context, the same as VIME. Therefore, we conduct experiments on VIME-R according to the methodology of Phase 2. We set the corruption rate, $p_m$, to 0.3 (the same as for vanilla VIME). 
% The number, $k$, of the k-nearest neighbors is set to \num{100} neighbors for both Bodmas and Ember datasets. 
Based on preliminary experiments with $k \in \{50, 100, 200, 500\}$ we found k = 100 to provide the best balance between neighborhood purity and corruption diversity. Smaller values produced overly conservative corruptions, while larger values diluted the locality signal. Therefore, the number, $k$, of the k-nearest neighbors is set to \num{100} neighbors for both Bodmas and Ember datasets.
We train VIME-R for 800 epochs, collect 16 evaluation checkpoints on each run, and perform five runs of the method on each dataset overall. Method evaluation also follows the methodology of Phase 2.

\begin{table}[t]
\centering
\caption{Neighbor purity analysis on the Ember dataset. For each value of $k$, we report the average number of same-class and opposite-class samples among the $k$ nearest neighbors, along with the purity (percentage of neighbors sharing the same class label).}
\label{tab:neighbor_purity}
\begin{tabular}{r rr c rr c}
\toprule
& \multicolumn{3}{c}{\textbf{Malware samples}} & \multicolumn{3}{c}{\textbf{Benign samples}} \\
\cmidrule(lr){2-4} \cmidrule(lr){5-7}
\textbf{$k$} & Same & Opp. & Purity & Same & Opp. & Purity \\
\midrule
1    & 1.00    & 0.00   & 100.0\% & 1.00    & 0.00   & 100.0\% \\
5    & 4.74    & 0.26   & 94.8\%  & 4.69    & 0.31   & 93.8\%  \\
10   & 9.31    & 0.69   & 93.1\%  & 9.20    & 0.80   & 92.0\%  \\
50   & 44.80   & 5.20   & 89.6\%  & 44.03   & 5.97   & 88.1\%  \\
100  & 88.03   & 11.97  & 88.0\%  & 86.40   & 13.60  & 86.4\%  \\
500  & 418.19  & 81.81  & 83.6\%  & 408.89  & 91.11  & 81.8\%  \\
1000 & 811.29  & 188.71 & 81.1\%  & 788.81  & 211.19 & 78.9\%  \\
2000 & 1574.83 & 425.17 & 78.7\%  & 1526.82 & 473.18 & 76.3\%  \\
\bottomrule
\end{tabular}
\end{table}

\subsubsection{Experimental Results}
% \textbf{}\quad
The results of the VIME-R experiments are presented in Table \ref{tab:vimer}. In it, we report relevant Homogeneity metrics for all three baselines and for the best performing method of Phase 2, which is VIME. For brevity and purposes of the best comparison, we report the mean Homogeneity across the five experimental runs of the methods (on both class and family labels), together with the respective standard deviations. 

The results tell a clear story: VIME-R is consistently able to beat the best performing methods of Phase 2 and thus achieve new state-of-the-art results. VIME-R outperforms all baselines on both datasets. Moreover, it beats VIME in all Homogeneity metrics, improving over VIME by 2.7\% (class) and 3.0\% (family) on BODMAS, and 5.8\% (class) and 3.9\% (family) on EMBER. This constitutes a significant improvement over the state-of-the-art.

\textbf{VIME-R surpasses the supervised ceiling on Ember.}\quad Perhaps the most remarkable finding is that VIME-R achieves 77.48\% family Homogeneity on Ember, surpassing the supervised ceiling established by SimSiam (76.53\%) in Phase 1. This means that a purely unsupervised method, relying solely on geometric proximity in the feature space, can produce clustering representations that are \emph{more informative} than those derived from explicit family-label supervision. We attribute this to the locality bias introduced by neighborhood-restricted corruption: by ensuring that corrupted features remain within the local data manifold, VIME-R learns representations that preserve fine-grained cluster structure more effectively than global approaches, whether supervised or unsupervised.

\begin{table}[t]
\centering
\caption{Homogeneity analysis of VIME-R against other representations used in Phase 2.}
\label{tab:vimer}
% \resizebox{0.75\textwidth}{!}{%
\begin{tabular}{ll cc}
\toprule
& & \multicolumn{2}{c}{\textbf{Homogeneity (Mean of Means)}} \\
\cmidrule(lr){3-4}
\textbf{Dataset} & \textbf{Method} & Class-level & Family-level \\
\midrule
\multirow{5}{*}{BODMAS}
& PCA          & 80.25 $\pm$ 1.09 & 71.96 $\pm$ 0.79 \\
& Autoencoder  & 82.37 $\pm$ 0.31 & 74.60 $\pm$ 0.32 \\
& UMAP         & 82.34 $\pm$ 0.60 & 80.43 $\pm$ 0.28 \\
& VIME         & \underline{87.02} $\pm$ 0.62 & \underline{80.87} $\pm$ 0.22 \\
& VIME-R       & \textbf{89.77} $\pm$ 0.31 & \textbf{83.83} $\pm$ 0.54 \\
\midrule
\multirow{5}{*}{EMBER}
& PCA          & 58.02 $\pm$ 0.19 & 66.13 $\pm$ 0.24 \\
& Autoencoder  & 58.32 $\pm$ 0.30 & 67.07 $\pm$ 0.37 \\
& UMAP         & \underline{67.13} $\pm$ 0.32 & \underline{75.55} $\pm$ 0.12 \\
& VIME         & 66.04 $\pm$ 0.41 & 73.61 $\pm$ 0.50 \\
& VIME-R       & \textbf{71.81} $\pm$ 0.45 & \textbf{77.48} $\pm$ 0.48 \\
\bottomrule
\end{tabular}%
% }
\end{table}

\begin{tcolorbox}[
    boxsep=3pt,          % Space between text and box border
    left=5pt,            % Left padding
    right=5pt,           % Right padding
    top=5pt,             % Top padding
    bottom=5pt,          % Bottom padding
    colback=white!15!gray!7, % Light grey background
    colframe=black!125,   % Slightly darker grey frame
    arc=1mm,             % Rounded corners
    outer arc=1mm,       % Outer rounded corners
    boxrule=0.4pt,       % Thickness of the frame
    fonttitle=\bfseries, % Make title bold (if you use a title)
    % title={\textbf{Malware Classification Accuracy:}} % Example title
]
\textbf{VIME-R Finding:}
Retrieval-augmented corruption via local neighborhood sampling (VIME-R) consistently and significantly outperforms vanilla VIME and all other unsupervised methods on both datasets, achieving new state-of-the-art results for binary program clustering. On EMBER, VIME-R is the only TRL method to surpass the strong UMAP baseline and the even stronger supervised ceiling.

\end{tcolorbox}

% \section{Possible Ablation ?}
% Use BODMAS and try the pair creation strategies on other models which did not make it to the previous results perhaps.

\section{Discussion}
\label{sec:discussion}
\subsection{Interpretation and Implications of Findings}

The collective results from the conducted experiments on RQ1 and RQ2 indicate that both self-supervised and tabular representation learning models can be used for malware clustering (or, rather, binary program clustering). The SSL approaches used in Phase 1, which aimed to assess the quality of the learned representations of the most prominent vision-based SSL methods, demonstrated strong performance in the supervised pair-learning context. Moreover, \say{ceiling} for pair learning methods was established. 

If reasonable data augmentation strategies (capable of creating positive pairs) for the malware domain are discovered (applicable not just to PE samples but also to vectorized representations), this would catalyze the mass adoption of such SSL methods in the malware domain. This holds mainly for the \emph{purely unsupervised} use of SSL methods, where strong data augmentation is necessary. This is the way forward to getting closer to the established supervised pair-learning ceiling. Moreover, even the SSL methods utilizing \emph{supervised pair} learning could become more frequently utilized in the future in the malware domain, as the (MLP versus BYOL) result on Ember indicates that, in certain cases, the representations extracted from SSL methods can outperform representations from fully-supervised models. 

Our results validate the efficacy of self-supervised tabular models, which achieved state-of-the-art performance in the binary program clustering task, beating strong unsupervised baselines by around 0.5-8\% on class Homogeneity and 0.4-9\% on family Homogeneity, depending on the specific dataset and method combinations. Notably, while UMAP provided very competitive results on both of the tested datasets, VIME remains more viable for large-scale applications. VIME scales efficiently with increasing sample sizes, whereas UMAP’s computational overhead becomes prohibitive as the dataset expands, presenting a significant limitation for real-time or high-volume malware analysis.

The data augmentation approach applied to BYOL and SimSiam in Phase 2, based on VIME, points to a lack of data augmentations for tabular datasets. Even more so, effective data augmentations for malware domain features are also scarce. Augmentation approaches that could specifically create positive pairs in the malware domain (for both malicious and benign samples) are highly needed for SSL methods like BYOL and SimSiam, since both methods were unable to achieve clustering quality even at baseline levels of Autoencoder or PCA. 

The introduction of VIME-R further reinforces the importance of the corruption strategy in tabular representation learning. By restricting the sampling pool to a sample's $k$ nearest neighbors, VIME-R bridges the gap between fully global (VIME) and fully supervised (Phase 1) corruption strategies. The improvements of 2.7–5.8\% in class Homogeneity and 3.0–3.9\% in family Homogeneity over vanilla VIME demonstrate that even a simple locality-aware modification to the corruption process can yield substantial gains. 

Crucially, VIME-R is the only unsupervised TRL method to surpass UMAP on the Ember dataset, a baseline that proved surprisingly resilient throughout our experiments. This result suggests that incorporating structural information about the data manifold into the corruption process is a highly effective inductive bias for binary program clustering. Moreover, VIME-R achieves this without any label information, relying solely on geometric proximity in the feature space. This positions retrieval-augmented corruption as a promising middle ground between domain-agnostic corruptions (which may introduce too much noise) and supervised pair generation (which requires costly labeling). 

% Moreover, the differences in cluster occupation discovered in Section \ref{sec:experiments_analysis_of_differences} clearly show underlying variation in how malware clustering representations behave when trained under a supervised versus an unsupervised regime. 

% \begin{itemize}
%     \item probably the main takeaway is that improvements can be made, however it depends on the dataset, and the whole setup (i.e. any labeled data available for model selection?) or maybe also, how much data we have.. this seems to bring more benefit to situations when we have let's say "smaller" / mid-sized data
%     \item how far are we from the ceiling with VIME/SWitchtab ? 
%     % \item we potentially found the potential ceiling for Homogeneities on the datasets
%     \item BYOL beats the rest, VIME beats other TRLs and unsup. baselines
%     \item UMAP seems to be strong, at least on smaller data; it has disadvantage - not that scalable
%     \item MLP not always best
%     \item VIME corruption for BYOL and SIMSIAM
%     \item Cluster Occupation differences 
% \end{itemize}

\subsection{Limitations}
Our work's first limitation is the reliance on static features (Ember featureset) for both datasets. While static analysis is computationally efficient, it may fail to capture complex runtime behaviors that dynamic analysis would reveal. Furthermore, as the threat landscape evolves, specific static features can become obsolete as attackers develop new evasion techniques to mimic benign file structures.

The Ember (2018) and Bodmas (2020) datasets, while standard benchmarks in recent literature \cite{jurevckova2024classification,mocko2025clustering}, represent specific temporal snapshots. Given the rapid evolution of malware, these samples may not fully capture the most recent polymorphic or fileless threats. However, they remain the most robust public benchmarks available for evaluating large-scale representation learning in this domain.

To isolate the impact of feature representations on clustering quality, we restricted our downstream evaluation to the K-Means algorithm. While alternative clustering methods (e.g., HDBSCAN or GMM) exist, using a consistent baseline ensures that observed improvements are attributable to the SSL/TRL representations rather than the clustering logic. Finally, while we conducted extensive preliminary experiments, the hyperparameter space for loss-function coefficients (e.g., in Barlow Twins and VICReg) and corruption probabilities (in VIME, VIME-R, and SCARF) is vast; further optimization might yield marginal gains but requires significant additional time investment and computational resources.

\subsection{Future work}
% \begin{itemize}
%     \item utilization of dynamic feature datasets
%     \item create some hybrid method
%     \item create some NN-based method ?
%     \item (a real) semi-supervised scenario could be explored, with classification of several thousands of samples for example
%     \item try out malmixer stuff (i.e. domain-based data augmentations)
%     \item utilization of an even more complex sampling method like VAE/GAN
%     \item VIME corruption which treats more informative and less informative features differently? 
% \end{itemize}

Several promising directions exist for extending this research. First, incorporating dynamic analysis datasets would provide a more holistic view of model behavior, though this requires generating larger benchmarks to overcome current public data scarcity. Second, the hybrid VIME-R architecture can be refined by implementing adaptive neighborhood sizes, distance-weighted sampling, or domain-aware retrieval corruption to further close the gap to the supervised ceiling. Third, extending the framework into a semi-supervised learning (SSL) scenario would leverage its ability to handle small amounts of labeled data (5–10\%), closely mirroring real-world cybersecurity triage. Finally, future iterations could transition from sampling marginal distributions toward modeling the joint probability distribution of features (inspired by \cite{li2025malmixer}) to capture complex inter-feature dependencies.

\section{Conclusion}
\label{sec:conclusion}
Our study builds on recent developments in self-supervised and tabular representation learning. This is the first research work to investigate the use of SSL and TRL methods for binary program clustering (an extension of malware clustering). 

% Furthermore, it is also the first work that studies the applicability of popular SSL models, namely BYOL, SimSiam, BarlowTwins, and VICReg, for tabular data in general. 
% The study builds on strong unsupervised baselines established in previous research. 
% We conduct our devised experiments on two popular public malware benchmark datasets, Bodmas and Ember. 
% By building upon robust unsupervised baselines and reaching new state-of-the-art results for binary program clustering on Bodmas and Ember via the utilization of tabular representation learning, we have advanced the state-of-the-art in malware feature representation. 
We advance malware feature representation by achieving state-of-the-art results in binary program clustering through specialized tabular representation learning on the Bodmas and Ember datasets.
% We reproduce the baseline results from a recent research work and set new state-of-the-art results for binary program clustering on both Bodmas and Ember via the utilization of tabular representation learning. 
Moreover, we explicitly report Homogeneity for both class (malware/benign) and family labels (e.g., Emotet, Mirai, etc.), which was often not the case in past research. This shows the differences in how well the representations can distinguish between malware and benign software, as well as how they perform at distinguishing individual malware families.  

% By establishing robust unsupervised baselines on the Bodmas and Ember datasets, we have advanced the state-of-the-art in malware feature representation. Notably, our work introduces a granular evaluation framework by explicitly reporting Homogeneity for both class and family labels

Phase 1 investigates popular SSL methods and their efficacy in supervised pair learning. We establish state-of-the-art results in supervised pair learning for binary program clustering across both datasets and multiple SSL methods. BYOL achieves the best results among the SSL methods, achieving 99.42\% class Homogeneity and 84.92\% family Homogeneity on Bodmas, even beating the MLP classifier. This allowed us to establish a performance \say{ceiling} for binary program clustering. This benchmark enables a direct comparative analysis for novel data augmentation strategies, providing a baseline for their effectiveness in the malware domain. Furthermore, we find that not all of the popular SSL methods are created equal. BarlowTwins and VICReg significantly underperform BYOL and SimSiam.

% Phase 2 explores an even more challenging task: improving malware clustering results in a fully unsupervised way, leveraging unsupervised representation learning. 
% In this task, VIME achieves the best results and beats the state-of-the-art binary program clustering baseline, establishing a new state-of-the-art result. Even more surprisingly, it beats the MLP classifier (a supervised baseline) on family Homogeneity on Ember. SwitchTab also outperforms the baseline, demonstrating that TRL methods are state-of-the-art learners even in the malware domain. Since the VIME feature corruption process lies on the edge of corruption and augmentation, we use it for BYOL and SimSiam and assess their representation quality. This process proves not to be powerful enough to reach the level of unsupervised baselines; however, it also manages to keep Homogeneity on both class and family labels reasonably high. This confirms the strong potential of VIME's data corruption/augmentation process. It also points to the need for specialized data augmentation techniques for the malware domain (performed not just on PE samples but also on vectorized features, in cases when PE features are not available).

Phase 2 evaluates unsupervised binary program clustering using tabular and self-supervised representation learning. VIME established a new state-of-the-art, surpassing the leading binary program clustering baseline and, remarkably, outperforming a supervised MLP on family Homogeneity on Ember. SwitchTab similarly beat the baseline, confirming tabular representation learning as a top-tier approach for malware. While applying VIME’s feature corruption to BYOL and SimSiam did not match unsupervised baselines, it maintained reasonably high Homogeneity across class and family labels. This validates VIME’s corruption/augmentation process while highlighting a critical need for domain-specific data augmentation tailored to vectorized malware features.

Finally, we proposed and experimentally validated VIME-R, a modification of VIME by introducing a retrieval-augmented corruption approach. VIME-R learns by corrupting samples via empirical marginal distribution  only based on samples that are in their close local neighborhood. VIME-R managed to achieve new state-of-the-art results and do it convincingly, with 2.7\% to 5.8\% differences in Homogeneity over the second-best unsupervised method, VIME. Introduction of retrieval-based data augmentation is the first step towards supervised-level clustering quality for the unsupervised task of binary program clustering (and malware clustering).

% Finally, we analyze the cluster assignments produced by three distinct representations: a deep Autoencoder (800 epochs), VIME, and an MLP classifier. This analysis reveals the structural differences in how each model encodes malware features. We observe that the MLP classifier facilitates feature abstraction, leading to a consolidation of malware families into fewer, broader clusters. Conversely, VIME promotes high-fidelity specialization, frequently dispersing samples from a single malware family across a more granular cluster distribution. These results validate the efficacy of self-supervised and tabular representation learning in enhancing malware clustering, providing a robust framework for future threat analysis.

\section*{Acknowledgment}
Acknowledgments have been omitted for the double-blind peer-review process.

\section*{LLM Usage Statement}
% We have used the AI assistant for grammar checks and sentence structure improvements. We have not used AI assistants in the research process.
The authors used an AI assistant to improve readability and language during manuscript preparation. The authors reviewed and edited the output and take full responsibility for the content of the publication.

\bibliographystyle{IEEEtran}
\bibliography{biblio}

% \usepackage{booktabs}

% \usepackage{graphicx}
% \usepackage{booktabs}
% \usepackage{pdflscape}
% \usepackage{graphicx}

% Place this in your appendix
\begin{landscape}
\begin{table}[ht]
\centering
\caption{Clustering evaluation results (mean of maxes) across methods and datasets.}
\label{tab:clustering_results}
\resizebox{\linewidth}{!}{%
\begin{tabular}{ll cccccc}
\toprule
& & \multicolumn{2}{c}{\textbf{Homogeneity}} & \multicolumn{2}{c}{\textbf{Completeness}} & \multicolumn{2}{c}{\textbf{V-Measure}} \\
\cmidrule(lr){3-4} \cmidrule(lr){5-6} \cmidrule(lr){7-8}
\textbf{Method} & \textbf{Dataset} & Class & Family & Class & Family & Class & Family \\
\midrule
autoencoder & bodmas & 84.48 $\pm$ 0.32 & 76.51 $\pm$ 0.30 & 10.50 $\pm$ 0.04 & 32.96 $\pm$ 0.15 & 18.67 $\pm$ 0.07 & 46.06 $\pm$ 0.18 \\
simsiamvime & bodmas & 77.80 $\pm$ 1.61 & 75.25 $\pm$ 1.11 & 9.61 $\pm$ 0.21 & 32.42 $\pm$ 0.47 & 17.11 $\pm$ 0.36 & 45.25 $\pm$ 0.61 \\
vicreg & bodmas & 38.88 $\pm$ 53.23 & 28.85 $\pm$ 39.55 & 67.10 $\pm$ 45.05 & 77.19 $\pm$ 31.35 & 11.99 $\pm$ 16.43 & 21.28 $\pm$ 29.22 \\
vime & bodmas & 87.85 $\pm$ 0.66 & 81.66 $\pm$ 0.37 & 10.74 $\pm$ 0.15 & 34.78 $\pm$ 0.37 & 19.13 $\pm$ 0.26 & 48.78 $\pm$ 0.43 \\
vime-r & bodmas & 90.83 $\pm$ 0.37 & 84.77 $\pm$ 0.65 & 10.59 $\pm$ 0.03 & 34.43 $\pm$ 0.13 & 18.96 $\pm$ 0.06 & 48.97 $\pm$ 0.23 \\
byolvime & bodmas & 73.11 $\pm$ 3.91 & 71.71 $\pm$ 2.53 & 9.27 $\pm$ 0.59 & 31.71 $\pm$ 1.03 & 16.46 $\pm$ 1.03 & 43.90 $\pm$ 1.35 \\
umap & bodmas & 82.34 $\pm$ 0.60 & 80.43 $\pm$ 0.28 & 9.93 $\pm$ 0.09 & 33.83 $\pm$ 0.21 & 17.73 $\pm$ 0.16 & 47.63 $\pm$ 0.25 \\
simsiam & bodmas & 98.46 $\pm$ 1.05 & 80.86 $\pm$ 2.01 & 12.54 $\pm$ 0.24 & 35.24 $\pm$ 0.69 & 22.24 $\pm$ 0.39 & 49.01 $\pm$ 1.07 \\
pca & bodmas & 80.25 $\pm$ 1.09 & 71.96 $\pm$ 0.79 & 10.23 $\pm$ 0.12 & 31.97 $\pm$ 0.26 & 18.14 $\pm$ 0.21 & 44.28 $\pm$ 0.40 \\
switchtab & bodmas & 84.17 $\pm$ 0.41 & 79.78 $\pm$ 0.47 & 10.11 $\pm$ 0.04 & 33.50 $\pm$ 0.14 & 18.04 $\pm$ 0.07 & 47.17 $\pm$ 0.19 \\
scarf & bodmas & 85.50 $\pm$ 0.28 & 76.93 $\pm$ 0.15 & 10.54 $\pm$ 0.03 & 33.07 $\pm$ 0.04 & 18.77 $\pm$ 0.06 & 46.24 $\pm$ 0.06 \\
mlp & bodmas & 97.94 $\pm$ 0.53 & 87.85 $\pm$ 1.18 & 11.89 $\pm$ 0.15 & 37.18 $\pm$ 0.36 & 21.20 $\pm$ 0.25 & 52.25 $\pm$ 0.49 \\
barlowtwins & bodmas & 77.40 $\pm$ 15.17 & 68.39 $\pm$ 13.43 & 9.33 $\pm$ 1.92 & 29.10 $\pm$ 6.21 & 16.65 $\pm$ 3.39 & 40.69 $\pm$ 8.36 \\
byol & bodmas & 99.42 $\pm$ 0.32 & 84.92 $\pm$ 1.61 & 12.25 $\pm$ 0.20 & 35.87 $\pm$ 0.40 & 21.79 $\pm$ 0.32 & 50.41 $\pm$ 0.59 \\
\midrule
autoencoder & ember & 58.80 $\pm$ 0.12 & 67.44 $\pm$ 0.23 & 6.10 $\pm$ 0.02 & 31.02 $\pm$ 0.11 & 11.05 $\pm$ 0.03 & 42.49 $\pm$ 0.14 \\
switchtab & ember & 65.29 $\pm$ 0.67 & 72.57 $\pm$ 0.24 & 6.43 $\pm$ 0.04 & 31.67 $\pm$ 0.17 & 11.70 $\pm$ 0.08 & 44.06 $\pm$ 0.14 \\
umap & ember & 67.13 $\pm$ 0.32 & 75.55 $\pm$ 0.12 & 6.20 $\pm$ 0.02 & 30.98 $\pm$ 0.05 & 11.36 $\pm$ 0.05 & 43.94 $\pm$ 0.06 \\
scarf & ember & 60.82 $\pm$ 0.06 & 69.45 $\pm$ 0.05 & 6.29 $\pm$ 0.01 & 31.91 $\pm$ 0.05 & 11.40 $\pm$ 0.01 & 43.72 $\pm$ 0.04 \\
simsiamvime & ember & 56.41 $\pm$ 0.92 & 68.86 $\pm$ 0.87 & 5.59 $\pm$ 0.07 & 30.44 $\pm$ 0.06 & 10.17 $\pm$ 0.12 & 42.16 $\pm$ 0.19 \\
vicreg & ember & 42.54 $\pm$ 16.23 & 47.63 $\pm$ 12.10 & 26.17 $\pm$ 41.27 & 52.09 $\pm$ 27.88 & 12.85 $\pm$ 2.12 & 37.54 $\pm$ 4.73 \\
pca & ember & 58.02 $\pm$ 0.19 & 66.13 $\pm$ 0.24 & 6.08 $\pm$ 0.01 & 30.73 $\pm$ 0.05 & 11.00 $\pm$ 0.02 & 41.96 $\pm$ 0.10 \\
vime & ember & 66.80 $\pm$ 0.38 & 74.36 $\pm$ 0.32 & 6.74 $\pm$ 0.07 & 33.31 $\pm$ 0.24 & 12.23 $\pm$ 0.11 & 45.81 $\pm$ 0.18 \\
barlowtwins & ember & 60.97 $\pm$ 2.96 & 61.41 $\pm$ 5.21 & 6.03 $\pm$ 0.26 & 26.83 $\pm$ 1.96 & 10.98 $\pm$ 0.47 & 37.33 $\pm$ 2.87 \\
vime-r & ember & 72.43 $\pm$ 0.36 & 77.79 $\pm$ 0.53 & 6.90 $\pm$ 0.02 & 32.86 $\pm$ 0.15 & 12.60 $\pm$ 0.05 & 46.20 $\pm$ 0.23 \\
simsiam & ember & 90.40 $\pm$ 2.28 & 76.53 $\pm$ 2.20 & 9.01 $\pm$ 0.15 & 33.83 $\pm$ 0.73 & 16.38 $\pm$ 0.29 & 46.92 $\pm$ 1.10 \\
byol & ember & 91.36 $\pm$ 2.64 & 75.87 $\pm$ 0.60 & 8.88 $\pm$ 0.26 & 32.53 $\pm$ 0.26 & 16.18 $\pm$ 0.48 & 45.53 $\pm$ 0.34 \\
mlp & ember & 96.09 $\pm$ 0.54 & 66.01 $\pm$ 6.86 & 9.93 $\pm$ 0.59 & 30.18 $\pm$ 2.48 & 18.00 $\pm$ 0.96 & 41.40 $\pm$ 3.57 \\
byolvime & ember & 52.92 $\pm$ 1.63 & 66.24 $\pm$ 1.28 & 5.39 $\pm$ 0.14 & 29.82 $\pm$ 0.29 & 9.78 $\pm$ 0.27 & 41.02 $\pm$ 0.45 \\
\bottomrule
\end{tabular}
}
\end{table}
\end{landscape}

\end{document}